\documentclass[%
amsmath,amssymb,floatfix,
reprint,%
twocolumn,%
longbibliography,
superscriptaddress,
]{revtex4-2}

\usepackage{amsmath,amsthm,latexsym,amssymb,amsfonts,epsfig}

\usepackage[utf8]{inputenc}
\usepackage{graphicx, texdraw}
\usepackage{bm}

\usepackage[normalem]{ulem} 

\usepackage{type1ec}
\usepackage{mathtools}
\usepackage{mathrsfs}

\usepackage{multirow}

\usepackage[outer=1.35cm, inner = 1.35cm, top= 1.91cm, bottom = 1.91cm]{geometry}  

\usepackage[tracking=true]{microtype}

\usepackage{enumerate}

\usepackage[colorlinks=true,linkcolor=MidnightBlue,urlcolor=MidnightBlue,citecolor=MidnightBlue,anchorcolor=MidnightBlue]{hyperref}

\usepackage{csquotes}

\usepackage{color}
\usepackage[dvipsnames]{xcolor}

\usepackage{amsmath}
\usepackage{amsfonts}
\usepackage{amssymb}

\newcommand{\vect}[1]{\boldsymbol{#1}}

\usepackage{ulem}

\usepackage{setspace}
\emergencystretch=\maxdimen
\allowdisplaybreaks

\definecolor{airforceblue}{rgb}{0.26, 0.54, 0.76}

\begin{document}

\title{Axisymmetric monopole and dipole flow singularities in proximity of a stationary no-slip plate immersed in a Brinkman fluid}

\author{Abdallah Daddi-Moussa-Ider} 
\email{abdallah.daddi-moussa-ider@ds.mpg.de}
\affiliation{Max Planck Institute for Dynamics and Self-Organization (MPI-DS), Am Fa{\ss}berg 17, 37077 G\"{o}ttingen, Germany}

\author{Yuto Hosaka} 
\affiliation{Max Planck Institute for Dynamics and Self-Organization (MPI-DS), Am Fa{\ss}berg 17, 37077 G\"{o}ttingen, Germany}

\author{Andrej Vilfan} 
\affiliation{Max Planck Institute for Dynamics and Self-Organization (MPI-DS), Am Fa{\ss}berg 17, 37077 G\"{o}ttingen, Germany}
\affiliation{Jo\v{z}ef Stefan Institute, 1000 Ljubljana, Slovenia}
\date{\today}

\author{Ramin Golestanian}
\affiliation{Max Planck Institute for Dynamics and Self-Organization (MPI-DS), Am Fa{\ss}berg 17, 37077 G\"{o}ttingen, Germany}
\affiliation{Institute for the Dynamics of Complex Systems, University of G\"{o}ttingen, 37077 G\"{o}ttingen, Germany}
\affiliation{Rudolf Peierls Centre for Theoretical Physics, University of Oxford, Oxford OX1 3PU, UK}

\date{\today}

\begin{abstract}

Green's function plays an important role in many areas of physical sciences and is a prime tool for solving diverse hydrodynamic equations in the linear regime. In the present contribution, the axisymmetric low-Reynolds-number Brinkman flow induced by monopole and dipole singularities in proximity of a stationary plate of circular shape is theoretically investigated. The flow singularities are directed along the central axis of the plate. No-slip boundary conditions are assumed to hold at the surface of the plate. The Green's functions are determined to a large extent analytically, reducing the solution of the linear hydrodynamic equations to well-behaved one-dimensional integrals amenable to numerical computation. In our approach, the Brinkman flow problem is formulated as a mixed boundary value problem that is subsequently mapped in the form of dual integral equations on the domain boundaries. Thereupon, the solution of the equations of fluid motion is eventually reduced to the solution of two independent Fredholm integral equations of the first kind. The overall flow structure and emerging eddy patterns are found to strongly depend on the magnitude of the relevant geometrical and physical parameters of the system. Moreover, the effect of the confining plate on the dynamics of externally driven or force-free particles is assessed through the calculation of the relevant hydrodynamic reaction functions. The effect of the plate on the locomotory behavior of a self-propelling active dipole swimmer is shown to be maximum when the radius of the plate is comparable to the distance separating the swimmer from the plate. Our results may prove useful for characterizing transport processes in microfluidic devices and may pave the way toward understanding and controlling of small-scale flows in porous media.

\end{abstract}

\maketitle

\section{Introduction}

Solutions of problems using Green's functions play an important role in many areas of physical sciences in which phenomena are modeled as linear processes~\cite{stakgold2011green}.
Knowledge of the Green's function associated with a linear differential equation is a building block for determining the solution of boundary value problems through the superposition principle~\cite{lanczos1996linear, greenberg2015applications}.
In modern fluid mechanics, Green's function is a prime tool in solving diverse hydrodynamic equations such as the linear Stokes equations governing the dynamics of viscous flows at low Reynolds numbers~\cite{happel12, kim13, pozrikidis1992boundary}.
Expressions of the Green's function of Stokes flow are documented for various types of confining boundary conditions~\cite{happel12}.

The Brinkman equation~\cite{brinkman1949permeability, brinkman1949permeability2} describes the effective flow of a fluid through a porous medium~\cite{ingham1998transport}.
The Brinkman description represents an extension of the classical form of Darcy's law~\cite{auriault2009domain} and was originally formulated to model the viscous flow past a sparse array of identical non-overlapping spherical particles~\cite{happel1958viscous, childress1972viscous, howells1974drag}.
In the context of biological fluid dynamics, the Brinkman description of porous media has been employed to predict the rate and pattern of growth of biofilm colonization within microfluidic chambers~\cite{kapellos2007hierarchical, valiei2012web, cogan2013pattern}, to examine the rheology of blood flow through the endothelial surface layer~\cite{damiano1996axisymmetric, secomb1998model, damiano1998effect, leiderman2008effects, weinbaum2003mechanotransduction}, or to investigate the formation of intravascular blood clots (thrombi) under viscous flow conditions~\cite{onasoga2013effect, leiderman2013influence}. 
In soft active matter research~\cite{Lauga2009, marchetti2013hydrodynamics, Bechinger2016}, it has been employed to describe a large variety of different phenomena, including the dynamical patterns inside metabolically active phase-separated enzymatic droplets \cite{testa2021Sustained} and the turbulent flows in active nematics with substrate friction \cite{Thampi2014}, among others.

In the Fourier domain, the Brinkman description is naturally linked to unsteady Stokes flow and linear viscoelastic fluids by a correspondence principle~\cite{feng2016boundary}. 
The fundamental solution for a viscous flow induced by an oscillating point force in the presence of a planar wall has been provided by Pozrikidis~\cite{pozrikidis1989singularity}, given later in a different form by Felderhof~\cite{felderhof05, felderhof2009flow}.
The latter noted the apparent analogy with the solution derived earlier by Sommerfeld to describe the propagation of waves in wireless telegraphy~\cite{sommerfeld1909ausbreitung} or the absorption of radial energy in dipole antennas~\cite{sommerfeld1942strahlungsenergie}.
More recently, a modified point-particle approximation and a refined method of reflections for unsteady Stokes flow near boundaries have been presented~\cite{simha2018unsteady}.
An overview summarizing the relevant literature and recent developments in the present context has been provided in Ref.~\onlinecite{fouxon2018fundamental}.
Meanwhile, the exact Fourier-space representation of Brinkman flow due to a regularized Brinkmanlet near a plane wall has been reported~\cite{nguyen2019computation}.

The fundamental solution for fluid flow in random porous media has been determined computationally using Stokesian dynamics simulations~\cite{durlofsky1987analysis} and compared with the Green's functions of the Brinkman equation. It has been demonstrated that the Brinkman equation describes the flow in porous media accurately only for volume fractions below 0.05. 
Accordingly, Brinkman description loses its predictive value for larger volume fractions, yet it can still describe the flow behavior in moderately concentrated porous media qualitatively.
In addition, a general computational method based on the boundary integral equation technique has been used to obtain the solutions to the Brinkman equation for the motion of a solid particle in the presence of planar confining boundaries~\cite{feng1998motion}.
Motivated by atomic force microscopy experiments, the small amplitude oscillation of a cantilever beam vibrating near a surface has further been studied under various conditions~\cite{green2005small, clarke2005drag, clarke2008three, clarke2006three}.

The Brinkman description has likewise been adopted to model two-dimensional (2D) fluid layers in contact with a solid substrate~\cite{evans1988, seki1993brownian, tserkovnyak2006conditions, ramachandran2010drag, ota2018three, hosaka2021hydrodynamic}.
Stokes' paradox~\cite{Landau1987}, which states that there are no bounded solutions for a 2D creeping flow around a disk, can be resolved in this way by incorporating a momentum decay.
On length scales much smaller than the hydrodynamic screening length~\cite{saffman1975, saffman1976, evans1988}, this phenomenological approach yields a logarithmic dependence that is consistent with the other models accounting for 3D fluids in the vicinity of a fluid layer~\cite{diamant2009hydrodynamic, ramachandran2011, oppenheimer2010correlated, hosaka2017lateral, hosaka2021nonreciprocal, hosaka2023hydrodynamics}.
Moreover, Blake and coworkers determined Green’s function of the Brinkman equation in a 2D fluid with anisotropic impermeability~\cite{kohr2008green}.
Meanwhile, the relation of 2D Brinkman flows to analogous Stokes flows has been highlighted~\cite{martin2019two}.
The 2D Brinkman model is particularly relevant for describing the dynamics of lipid bilayer membranes on solid substrates that mimic the generic role of the extracellular matrix~\cite{albertsbook, tanaka2005polymer}, or to describe the flow behavior in microfluidic systems in which the depth-wise dimension is small \cite{battat2019particle}.
The swimming behavior of flagellated model microswimmers in Brinkman fluids has further been examined both in 2D~\cite{leiderman2016swimming} and 3D~\cite{ho2019three} spaces.

In each of the above-mentioned theoretical and computational studies, the confining wall was assumed to be of infinite extent along the horizontal plane.
Here, we consider instead the hydrodynamic problem for monopole and dipole singularities acting near a no-slip plate of finite radius immersed in a Brinkman medium. 
The present contribution addresses this much less investigated configuration.
Indeed, in many biologically relevant applications, finite-size effects become of crucial importance for a reliable and accurate description of various fluid dynamics and transport processes spanning many length scales.

Here we report on a semi-analytical theory for the viscous flow resulting from a point force (Brinkmanlet) or dipole singularities in proximity of a circular no-slip disk. We formulate the flow problem as a standard mixed boundary value problem that we subsequently transform into two independent dual integral equations. 
We provide the solution of the resulting Fredholm integral equations in terms of convergent infinite series.
We demonstrate the emergence of toroidal eddies arising in the fluid domain and show that the overall flow structure depends on the magnitude of the porosity coefficient and the ratio between the height of the singularity above the plate and the radius of the plate.
We also probe the effect of finite size on the dynamics of an externally driven colloidal particle or a force-free microswimmer through the evaluation of the hydrodynamic reaction resulting from the monopole and dipole flow fields, respectively.

\section{Problem statement}

\begin{figure}
    \centering
    \includegraphics[width=0.7\columnwidth]{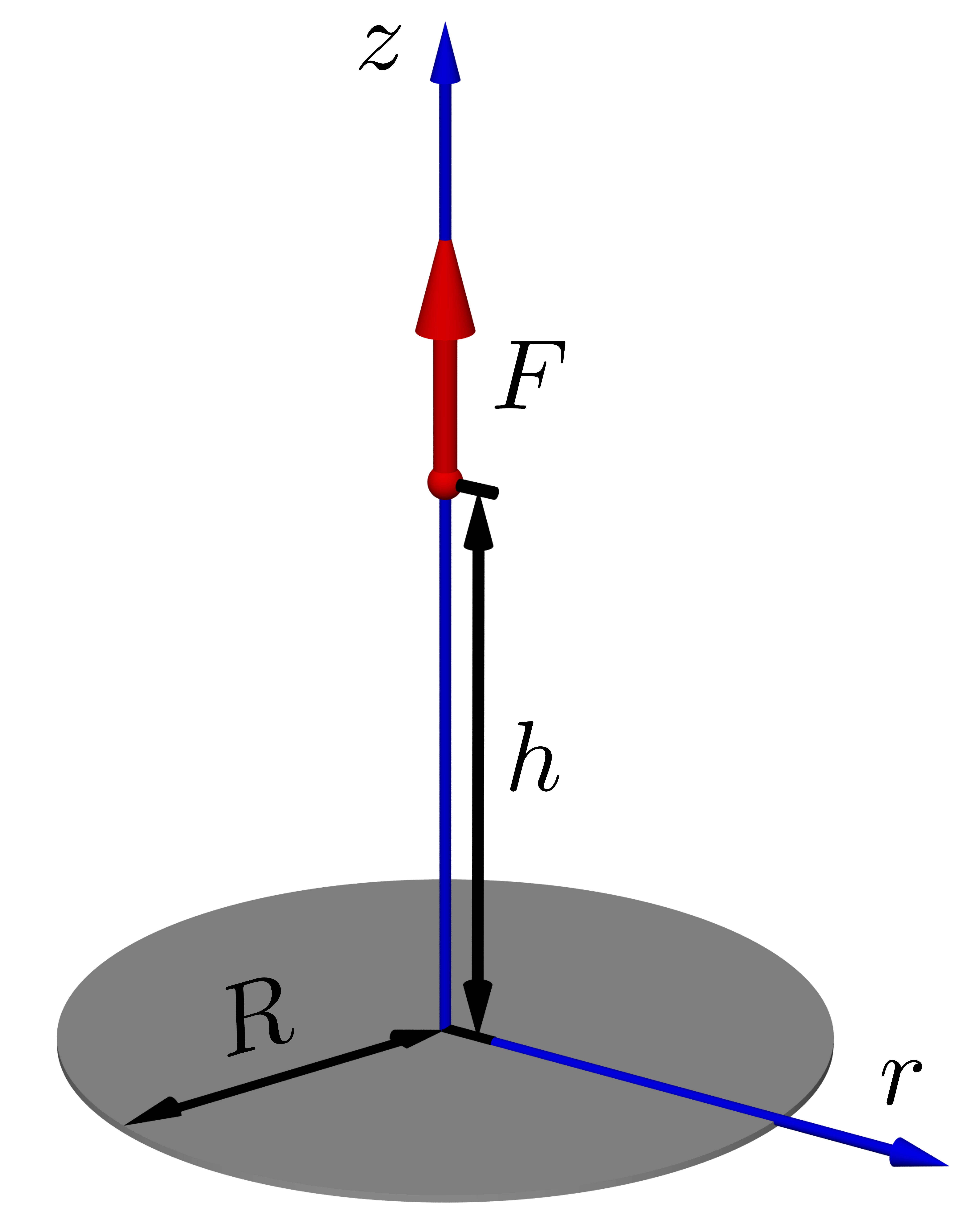}
    \caption{(Color online) Graphical illustration of the problem setup.
    A point force of strength~$F$ is located axisymmetrically at height~$h$ above a no-slip circular disk with radius~$R$ immersed in a Brinkman fluid medium.}
    \label{fig:intro}
\end{figure}

We examine the low-Reynolds-number flow induced by a point-force singularity acting on the symmetry axis of an immobile disk of radius~$R$ immersed in a Brinkman fluid medium. 
We denote by~$h$ the distance between the location of the point force and the center of the disk.
We assume the disk to be extended along the $xy$ plane.
Owing to the axisymmetric nature of the problem, we employ a system of cylindrical coordinates such that the point force~$\vect{F} = F \hat{\vect{e}}_z $ is acting along the vertical direction defined by the unit vector~$\hat{\vect{e}}_z$ normal to the plate; 
see Fig.~\ref{fig:intro} for an illustrative schematic of the system setup.

The flow is governed by the Brinkman equation and the incompressibility condition~\cite{brinkman1949permeability}
\begin{equation}
    -\boldsymbol{\nabla}p + \eta \boldsymbol{\nabla}^2 \vect{v} - \eta \alpha^2  \vect{v} + \vect{f}  = \vect{0} \, , \quad
    \boldsymbol{\nabla} \cdot \vect{v} = 0 \, , \label{Brinkman-Eqs}
\end{equation}
where $\eta$ is the dynamic viscosity of the Newtonian fluid, $\alpha^2$~is the impermeability of the porous medium, which has the dimension of (length)$^{-2}$, and $\vect{v}$ and~$p$ are the fluid velocity and pressure, respectively. 
Here, $\vect{f} = \vect{F} \delta \left( \vect{r} - h \hat{\vect{e}}_z \right)$ is a point-force density acting on the surrounding fluid above the plate at $(0,0,h)$.

Brinkman Equation can be projected and expressed in the system of axisymmetric cylindrical coordinates as
\begin{subequations} \label{eq:momentum}
\begin{align}
	-\frac{\partial p}{\partial r} + \eta \left( \frac{\partial}{\partial r} \left( \frac{1}{r} \frac{\partial }{\partial r} \left( r v_r \right) \right) + \frac{\partial^2 v_r}{\partial z^2} - \alpha^2 v_r \right) &= 0 \, , \\
	-\frac{\partial p}{\partial z} + \eta \left( \frac{1}{r} \frac{\partial}{\partial r} \left( r \, \frac{\partial v_z}{\partial r} \right) + \frac{\partial^2 v_z}{\partial z^2} - \alpha^2 v_z \right) + f &= 0 \, ,
\end{align}
\end{subequations}
wherein $f = F \delta \left(z-h\right)\delta(r)/\left(\pi r\right)$.
In addition, $v_r$ denotes the radial and~$v_z$ the axial component of the fluid velocity.
The flow incompressibility condition in cylindrical coordinates reads
\begin{equation}
\frac{1}{r} \frac{\partial}{\partial r} \left( r v_r \right) + \frac{\partial v_z}{\partial z} = 0 \, .
\label{eq:incompress}
\end{equation}

Equations~\eqref{eq:momentum} and \eqref{eq:incompress} are subject to the regularity condition of vanishing fields at infinity in addition to the boundary condition of no slip on the surface of the disk.
Specifically, 
\begin{equation}
    \vect{v} (r, z=0) = \vect{0} \qquad \text{for} \quad r < R \, .
\end{equation}
In addition, we require the natural continuity of the viscous traction in the flow region outside the disk for $r > R$.

Owing to the linearity of the governing equations of fluid motion, a solution can be obtained more conveniently using the image technique~\cite{kim13}.
In this approach, we express the solution for the hydrodynamic fields as a superposition of the known solution in an unbounded fluid medium, i.e., in the absence of the disk, and a particular solution of the homogeneous differential equation that is required to satisfy the boundary conditions prescribed on the surface of the confining disk.
Accordingly, the hydrodynamic fields can be expressed in the form
\begin{equation}
    \vect{v} = \frac{F}{8\pi\eta} \left( \vect{G}^\infty + \vect{G} \right)  \, , \qquad
    p = \frac{F}{4\pi} \left( P^\infty + P \right) \, ,
\end{equation}
wherein $\left\{ \vect{G}^\infty, P^\infty \right\}$ correspond to the Green's functions in infinite space and $\left\{ \vect{G}, P \right\}$ the contributions of the boundary. The Green's functions for the velocity have the dimensions of (length)$^{-1}$ and those for the pressure have the dimensions (length)$^{-2}$.

The velocity and pressure fields induced by a point-force singularity acting in an infinite Brinkman fluid, commonly known as the free-space Brinkmanlet, are given by~\cite{howells1974drag}
\begin{subequations} \label{eq:brinkmanlet}
\begin{align}
	G_r^\infty &= B_2 \,  \frac{r \left(z-h\right)}{s^3} \, , \\
	G_z^\infty &= \frac{B_1}{s} + \frac{B_2}{s^3} \left(z-h\right)^2 \, , \\
	P^\infty   &= \frac{z-h}{s^3} \, .
\end{align} 
\end{subequations}
We have defined $s = \left( r^2 + \left(z-h \right)^2 \right)^\frac{1}{2}$ as the distance from the singularity position, and $B_1$ and~$B_2$ are two functions of this distance defined as
\begin{subequations}
\begin{align}
	B_1 &= 2e^{-\alpha s} \left( 1 + \frac{1}{\alpha s} + \frac{1}{\alpha^2 s^2} \right) - \frac{2}{\alpha^2 s^2} \, , \\
	B_2 &= \frac{6}{\alpha^2 s^2} - 2 e^{-\alpha s} \left( 1 + \frac{3}{\alpha s} + \frac{3}{\alpha^2 s^2} \right) \, .
\end{align}
\end{subequations}
In particular, it can be checked that in the limit $\alpha \to 0$, $B_1 = B_2 = 1$, reducing Eqs.~\eqref{eq:brinkmanlet} into the classical Oseen tensor~\cite{oseen27} for a Newtonian fluid.

\section{Monopole flow}

Before delving further into the calculation details, we summarize the main mathematical steps leading to the image solution.
First, the governing equations of fluid motion are Hankel transformed, leading to ordinary differential equations in the Hankel transformed variables.
In this way, the image solution can be presented in the form of infinite integrals over the wavenumber upon inverse Hankel transform.
Secondly, applying the boundary conditions yields dual integral equations that are subsequently transformed into Fredholm integral equations of the first kind following the solution approach outlined by Sneddon and Copson.
Thirdly, the resulting integral equations are solved numerically by approximating the definite integrals as midpoint Riemann sums. 
Finally, the hydrodynamic fields can be computed numerically at every point in the fluid domain.

\subsection{Hankel-transformed equations}

For a given radial function~$\phi(r)$, $r \in [0, \infty)$, we use the convention~\cite{davies12book}
\begin{equation}
\widetilde{\phi} (q) =
\mathcal{H}_\nu \left\{ \phi(r) \right\} = 
\int_0^\infty r\phi(r) J_\nu (qr) \, \mathrm{d} r \, , 
\end{equation}
for the forward transform, and
\begin{equation}
\phi (r) =
\mathcal{H}_\nu^{-1} \left\{ \widetilde{\phi}(q) \right\} = 
\int_0^\infty q \widetilde{\phi}(q) J_\nu (qr) \, \mathrm{d} q \, , 
\end{equation}
for the inverse transform. 
Here, $J_\nu$ is the Bessel function of the first kind of order $\nu \ge -\frac{1}{2}$.
We use the first-order Hankel transform for the radial component of the fluid velocity and zeroth-order Hankel transform for the axial velocity and pressure.
Defining $\widetilde{G}_r = \mathcal{H}_1 \left\{ G_r \right\}$, $\widetilde{G}_z = \mathcal{H}_0 \left\{ G_z \right\}$, and $\widetilde{P} = \mathcal{H}_0 \left\{ P \right\}$, the resulting Hankel transforms of the homogeneous momentum equations~\eqref{eq:momentum} read
\begin{subequations}
\begin{align}
	q \widetilde{P} + \frac{1}{2} \left( \frac{\partial^2 }{\partial z^2} - \left( q^2+\alpha^2 \right) \right) \widetilde{G}_r &= 0 \, , \\
	- \, \frac{\partial \widetilde{P}}{\partial z} + \frac{1}{2} \left( \frac{\partial^2 }{\partial z^2} - \left( q^2+\alpha^2 \right) \right) \widetilde{G}_z &= 0 \, .
\end{align}%
\label{eq:momentum-hankel}%
\end{subequations}
The incompressibility equation \eqref{eq:incompress} can likewise be Hankel transformed as
\begin{equation}
q \widetilde{G}_r + \frac{\partial \widetilde{G}_z}{\partial z} = 0 \, . \label{eq:incompress-hankel}
\end{equation}

Accordingly, the partial differential equations governing the fluid motion have now been transformed into a system of ordinary differential equations in the Hankel transformed variables.

It follows from Eq.~\eqref{eq:incompress-hankel} that $\widetilde{G}_r = - \left( 1/q \right) \left( \partial \widetilde{G}_z / \partial z \right) $.
By combining Eqs.~\eqref{eq:momentum-hankel} so as to eliminate the pressure field, a fourth-order differential equation for the normal component of the fluid velocity is obtained, namely,
\begin{equation}
\left(
\frac{\partial^4 }{\partial z^4} - \left(2q^2+\alpha^2\right) \frac{\partial^2}{\partial z^2} + q^2 \left( q^2+\alpha^2 \right) \right) \widetilde{G}_z = 0 \, .
\label{eq:fourth-order}
\end{equation}

Considering the regularity condition as $z \to \pm \infty$, the general solution of Eq.~\eqref{eq:fourth-order} is given by
\begin{equation}
\widetilde{G}_z^\pm = c_1^\pm   e^{\mp qz} + c_2^\pm  e^{\mp Qz} \, ,
\end{equation}
wherein $Q = \left( q^2+\alpha^2 \right)^\frac{1}{2}$, and $c_i^\pm$, $i = 1,2$, are wavenumber-dependent functions, to be subsequently determined from the underlying boundary conditions.
Plus and minus signs in superscripts refer to the flow variables above the plate $(z>0)$ and below the plate $(z<0)$, respectively. 
The solution for the radial velocity reads
\begin{equation}
\widetilde{G}_r^\pm = \pm c_1^\pm  e^{\mp qz} \pm \frac{Q}{q} \, c_2^\pm  \, e^{\mp Qz} \, .
\end{equation}
The corresponding solution for the pressure field reads
\begin{equation}
\widetilde{P}^\pm = \pm \frac{\alpha^2}{2q} \, c_1^\pm  e^{\mp qz}  \, .
\end{equation}

The solutions of the hydrodynamic fields in real space are readily obtained upon inverse Hankel transform as $G_r = \mathcal{H}_1^{-1} \left\{ \widetilde{G}_r \right\}$, $G_z = \mathcal{H}_0^{-1} \left\{ \widetilde{G}_z \right\}$, and $P = \mathcal{H}_0^{-1} \left\{ \widetilde{P} \right\}$.
Ultimately, the image velocity and pressure fields can be expressed in integral forms over the wavenumber~$q$ as
\begin{subequations} \label{eq:image_final}
\begin{align}
    G_r^\pm &= \pm \int_0^\infty \left( q c_1^\pm  e^{-q|z|} + Q c_2^\pm   e^{-Q|z|}  \right) J_1 (qr) \, \mathrm{d} q \, , \\
    G_z^\pm &= \int_0^\infty q \left( c_1^\pm  e^{-q|z|} + c_2^\pm   e^{-Q|z|} \right) J_0 (qr) \, \mathrm{d}q \, , \label{vzP} \\
    P^\pm &= \pm \frac{\alpha^2}{2} \int_0^\infty c_1^\pm  e^{-q|z|} J_0 (qr) \, \mathrm{d} q \, .
\end{align}
\end{subequations}

\subsection{Dual integral equations}

Having derived a general integral representation of the image solution for an axisymmetric Brinkman flow field, we next determine the wavenumber-dependent quantities~$c_i^\pm$, $i = 1,2$, such that the underlying boundary conditions imposed at $z=0$ are satisfied.

It follows from the continuity of radial and axial components of the fluid velocity that
\begin{subequations} \label{BCs:continuity}
\begin{align}
	c_1^+ -c_1^-  +c_2^+  -c_2^-   &= 0 \, , \\
	q \left( c_1^+ +c_1^-   \right) + Q \left( c_2^+  +c_2^-   \right) &= 0 \, .
\end{align}
\end{subequations}

On the one hand, the equations for the inner problem are obtained by imposing the no-slip boundary condition at the surface of the disk as 
\begin{subequations} \label{BCs:no-slip}
\begin{align}
	\int_0^\infty \left( q c_1^+  + Q c_2^+   \right) J_1(qr) \, \mathrm{d}q  &= -G_r^\infty (r, z=0) \, , \\
	\int_0^\infty q \left( c_1^+ +c_2^+   \right) J_0 (qr) \, \mathrm{d} q 
	 &= -G_z^\infty (r, z = 0) \, ,
\end{align}
\end{subequations}
for $r < R$.
On the other hand, the equations for the outer problem are obtained by requiring continuity of the radial and axial components of the viscous traction.
Then,
\begin{subequations} \label{outer-problem}
\begin{align}
	\int_0^\infty A(q) J_1 (qr) \, \mathrm{d} q &= 0 \, , \\
	\int_0^\infty B(q) J_0 (qr) \, \mathrm{d} q &= 0 \, ,
\end{align}
\end{subequations}
for $r>R$.
The wavenumber-dependent integrands are given by 
\begin{subequations}  \label{AB}
\begin{align}
    A &= 2q^2\left( c_1^+ -c_1^-  +c_2^+  -c_2^-   \right) + \alpha^2 \left( c_2^+  -c_2^-   \right) \, , \\
    B &= 2qQ \left( c_2^+  +c_2^-   \right) + \left( 2q^2+\alpha^2 \right) \left( c_1^+ +c_1^- \right) \, .
\end{align}
\end{subequations}
Moreover, solving Eqs.~\eqref{BCs:continuity} for the coefficients $c_2^+$ and~$c_2^-$ associated with the lower fluid domain $(z<0)$ yields
\begin{subequations} \label{c3c4}
\begin{align}
	c_2^+   &= \frac{1}{2} \left( c_1^-   - c_1^+  \right) 
	-\frac{q}{2Q} \left( c_1^+  + c_1^-   \right) \, , \\
	c_2^-   &= \frac{1}{2} \left( c_1^+  - c_1^-   \right)
	- \frac{q}{2Q} \left( c_1^+  + c_1^-   \right) \, .
\end{align}
\end{subequations}

It follows from Eqs.~\eqref{AB} and~\eqref{c3c4} that $c_1^\pm  = \left( B \pm A\right)/ \left( 2\alpha^2 \right)$ and $c_2^\pm = \left(\mp A-qB/Q\right)/ \left( 2\alpha^2 \right)$.
Upon substitution of these forms into Eqs.~\eqref{BCs:no-slip}, the integral equations for the inner problem are obtained as
\begin{subequations} \label{inner-problem}
\begin{align}
	\int_0^\infty \frac{2}{\alpha^2} \left( Q-q \right) A(q) J_1(qr) \, \mathrm{d} q &= -4 G_r^\infty (r, z=0) \, , \\
	\int_0^\infty \frac{2q}{\alpha^2} \left( 1 - \frac{q}{Q} \right) B(q) J_0(qr) \, \mathrm{d} q &= -4 G_z^\infty (r, z=0) \, ,
\end{align}
\end{subequations}
for $r<R$.
The flow problem is thus reduced to the dual integral equations given by Eqs.~\eqref{outer-problem} and~\eqref{inner-problem} for the unknown wavenumber-dependent functions $A(q)$ and~$B(q)$.

A standard approach to derive solutions of the resulting integral equations can be achieved by means of the theory of Mellin transforms~\cite{titchmarsh48book, tranter51}.
Here, we prefer to follow an alternative route based on the well-established solution recipes described by Sneddon~\cite{sneddon60, sneddon66} and Copson~\cite{copson1947problem, copson61} for solving dual integral equations.
This solution approach has been employed by some of us to solve various Stokes flow problems involving finite-sized boundaries. 
These include the theoretical investigation of the viscous flow induced by diverse types of force or source singularities acting in proximity of a deformable disk possessing shear and bending~\cite{daddi19jpsj, daddi2020asymmetric}, near a no-slip circular disk~\cite{daddi2020dynamics, daddi2021steady, daddi2022diffusiophoretic}, or between two parallel coaxially positioned no-slip disks of equal size~\cite{daddi2020axisymmetric, daddi2022stokeslet}.
Accordingly, the problem is reduced to the search for the solution of Fredholm integral equations, which can, under special circumstances, also be inverted analytically for simple boundary conditions.

\subsection{Formulation of solution}

To satisfy the integral equations for the outer problem stated by Eqs.~\eqref{outer-problem}, we express the solutions in terms of \textit{finite} Fourier transforms of the forms 
\begin{subequations} \label{AB-solution-form}
\begin{align}
	A(q) &= -4 q \int_0^R f(t) \sin (qt) \, \mathrm{d} t \, , \\
	B(q) &= -4 q\int_0^R g(t) \cos (qt) \, \mathrm{d} t \, .
\end{align}
\end{subequations}
We will now demonstrate that these expressions fulfil the condition imposed in the outer domain.
In fact, using integration by parts, we have
\begin{subequations}
\begin{align}
	\frac{A(q)}{4} &= \big. f(t) \cos \left(qt\right) \big|_{t=0}^{t=R}
	- \int_0^R f'(t) \cos (qt) \, \mathrm{d} t \, ,  \\[2pt]
	\frac{B(q)}{4} &= \int_0^\infty g'(t) \sin (qt) \, \mathrm{d}t -g(R)\sin \left(qR\right)  .
\end{align}
\end{subequations}

Since~\cite{watson95}
\begin{subequations}
\begin{align}
	\int_0^\infty \cos(qt) J_1(qr) \, \mathrm{d}q &=
	 \frac{1}{r} - \frac{t H(t-r)}{r\left( t^2-r^2 \right)^\frac{1}{2}} , \\
	\int_0^\infty \sin(qt) J_0(qr) \, \mathrm{d}q &=
	\frac{H(t-r)}{\left( t^2-r^2 \right)^\frac{1}{2}}\, ,
\end{align}
\end{subequations}
with $H$ representing the Heaviside step function, it can readily be checked that the integral equations for the outer problem are satisfied when $f(0)=0$.
We will show later on that indeed this is the case here.

Ultimately, by substituting the integral representations of $A(q)$ and~$B(q)$ as given by Eqs.~\eqref{AB-solution-form} into the equations for the inner problem stated by Eqs.~\eqref{inner-problem}, and invoking the expressions of the free-space Brinkmanlet stated by Eqs.~\eqref{eq:brinkmanlet}, the following integral equations for the unknown functions $f(t)$ and~$g(t)$ are obtained.
Specifically,
\begin{subequations} \label{eq:dual-final}
\begin{align}
	\int_0^R f(t) \Gamma_1 (r,t) \, \mathrm{d}t &= \beta_2 \, \frac{hr}{\rho^3}  \, , \\[2pt]
	\int_0^R g(t) \Gamma_2 (r,t) \, \mathrm{d}t &= \frac{1}{\rho} \left( \beta_1 + \beta_2 \, \left( \frac{h}{\rho} \right)^2 \right) \, ,
\end{align}
\end{subequations}
for $r<R$, wherein $\rho = \left( r^2+h^2 \right)^\frac{1}{2}$.
In addition, we have defined the abbreviations $\left. \beta_i = B_i \right|_{s = \rho}$, $i = 1, 2$.
Moreover, the kernel functions $\Gamma_1$ and~$\Gamma_2$ are expressed in term of infinite integrals over the wavenumber as
\begin{subequations} \label{Gammms}
\begin{align}
	\Gamma_1 &= \frac{2}{\alpha^2} \int_0^\infty
	q \left( \left( q^2+\alpha^2 \right)^\frac{1}{2} - q \right)
	\sin(qt) J_1(qr) \, \mathrm{d} q \, , \label{Gamma1} \\
	\Gamma_2 &= \frac{2}{\alpha^2} \int_0^\infty
	q^2 \, \frac{\left(q^2+\alpha^2\right)^\frac{1}{2} - q}{\left(q^2+\alpha^2\right)^\frac{1}{2}} \, \cos(qt) J_0(qr) \, \mathrm{d} q \, . \label{Gamma2}
\end{align}
\end{subequations}
Equations~\eqref{eq:dual-final} are Fredholm integral equations of the first kind for the unknown functions $f(t)$ and~$g(t)$ defined on the interval $[0,R]$.

In the Stokes flow limit for which $\alpha \to 0$, exact analytical solutions of Eqs.~\eqref{eq:dual-final} can be obtained.
These have been obtained by Kim~\cite{kim83} earlier using an analogous dual integral equation approach.
However, for the general case of arbitrary values of~$\alpha$, an analytical evaluation of the infinite integrals defining the kernel functions given in by Eqs.~\eqref{Gammms} is rather delicate and far from being trivial.
Even though a direct analytical evaluation of the integral is seemingly impossible, we will show in the following that both infinite integrals can conveniently be transformed into convergent series functions amenable to numerical evaluation upon truncation.
The chief benefit of our approach is that the solution of the axisymmetric Brinkman flow problem is reduced to one dimensional numerical integration over a bounded interval.

\subsection{Evaluation of the kernel functions}

The core idea of our semi-analytical approach consists of expressing the Bessel functions in Eqs.~\eqref{Gammms} in terms of the integral form given by Poisson [see e.g., Gröbner and Hofreiter~\cite{grobner1966integraltafel}]
\begin{equation}
	J_\nu(z) = \frac{2 \left( \frac{z}{2} \right)^\nu}{\pi^\frac{1}{2} \Gamma \left( \nu + \frac{1}{2} \right)} 
	\int_0^1 \left( 1-w^2 \right)^{\nu-\frac{1}{2}} \cos (zw) \, \mathrm{d}w \, ,
\end{equation}
and interchanging the order of integration with respect to the variables~$w$ and~$q$.
Here, $\Gamma$~denotes the Euler Gamma function~\cite{abramowitz72}.
Performing integration first with respect to the wavenumber~$q$ between 0 and $\infty$ yields a definite integral with respect to the variable $w$ between 0 and 1.
By using the series expansion representation of the integrands, an analytical evaluation of the resulting definite integrals is possible.
Correspondingly, the kernel functions can be cast in the form
\begin{subequations} \label{eq:kernels_final}
	\begin{align}
		\Gamma_1 &= \Gamma_1^0 + \alpha^2 r \left( \frac{\alpha t}{15} + 
		\Psi_1 +  2 \, \Psi_2 H (r-t) \right) \, , \\
		\Gamma_2 &= \Gamma_2^0 - \alpha \left( \frac{4}{3} 
		-\Psi_3  + 2 \Psi_4  H(r-t) \right) \, , 
	\end{align}
\end{subequations}
wherein $\Gamma_i^0$, for $i =1,2$, stand for the kernels in the Stokes flow limit as $\alpha \to 0$, given by
\begin{equation}
\Gamma_1^0 (r,t) = \frac{t H(r-t)}{r \left( r^2-t^2\right)^\frac{1}{2}} \, , \qquad
\Gamma_2^0 (r,t) = \frac{H(r-t)}{\left( r^2-t^2\right)^\frac{1}{2}}\, .
\end{equation}
In addition,
\begin{subequations}
	\begin{align}
		\Psi_1 (r,t) &= \int_0^1 
		\hat{\Psi}_1 (w) \left( 1-w^2 \right)^\frac{1}{2} \mathrm{d}w \, , \label{psi1} \\
		\Psi_2 (r,t) &= \int_{\frac{t}{r}}^1 \hat{\Psi}_2 (w)
		\left( 1-w^2 \right)^\frac{1}{2} \, \mathrm{d}w \, , \label{psi2} \\
		\Psi_3 (r,t) &= \int_0^1 \hat{\Psi}_3 (w) \left( 1-w^2 \right)^{-\frac{1}{2}} \mathrm{d} w \, , \label{psi3}\\
		\Psi_4 (r,t) &= \int_{\frac{t}{r}}^1 \hat{\Psi}_4 (w) \left( 1-w^2\right)^{-\frac{1}{2}} \mathrm{d}w \, , \label{psi4}	
	\end{align}
\end{subequations}
with the integrands given by
\begin{subequations}
	\begin{align}
		\hat{\Psi}_1 (w) &= \frac{\Phi_2(w_-)}{w_-^2} + \frac{\Phi_2(w_+)}{w_+^2} + \frac{\Phi_3(w_-)}{w_-} + 
		\frac{\Phi_3(w_+)}{w_+} \, , \\
		\hat{\Psi}_2 (w) &= \frac{I_2(w_-)}{w_-^2} + \frac{I_3(w_-)}{w_-} \, ,  \\
		\hat{\Psi}_3 (w) &= \Phi_1(w_-) + \Phi_1(w_+) + \frac{\Phi_4 (w_-)}{w_-} + \frac{\Phi_4(w_+)}{w_+} \, , \\
		\hat{\Psi}_4 (w) &= \frac{3}{w_-} \, I_2(w_-) + I_3(w_-) \, .
	\end{align}
\end{subequations}
Here, we have defined the abbreviation $w_\pm = \alpha \left( t \pm r w \right)$.
In addition, $\Phi_1(z) = I_3(z)-L_1(z)$, $\Phi_2 (z) = L_2(z) - I_2(z)$, $\Phi_3 (z) = L_3(z) - I_3(z)$, and $\Phi_4(z) = 3I_2(z)+L_2(z)$, $z \in \mathbb{C}$.
Here, $L_n$ and~$I_n$ represent the $n$th order modified Struve function and modified Bessel function of the first kind, respectively.

To be able to make analytical progress, we use the series representation of the modified Struve function and modified Bessel function of the first kind, given by~\cite{abramowitz72}
\begin{subequations} \label{StruveL-Bessel-Expansions}
	\begin{align}
		L_\nu(z) &= \sum_{m \ge 0} \frac{\left( \frac{z}{2} \right)^{2m+\nu+1}}{\Gamma \left( m+\frac{3}{2}\right) \Gamma \left( m+\nu+\frac{3}{2} \right)} \, , \\
		I_\nu(z) &= \sum_{m \ge 0} \frac{\left( \frac{z}{2} \right)^{2m+\nu}}{\Gamma \left( m+1\right) \Gamma \left( m+\nu+1 \right)} \, .
	\end{align}
\end{subequations}
We note that the Gamma function has the recursive property $\Gamma(z+1)=z\Gamma(z)$ and that for a positive integer $n$ we have $\Gamma(n+1) = n!$.

Thus, by making use of the series expansions given by Eqs.~\eqref{StruveL-Bessel-Expansions}, and defining the dimensionless quantities $\sigma = \alpha t/2$ and $\mu = r/t$, the functions $\Psi_1$ and~$\Psi_2$ defining the kernel $\Gamma_1$ are obtained as
\begin{subequations} \label{Psi1_Psi2}
    \begin{align}
    \Psi_1 &= - \frac{\alpha t}{15} + \sum_{m \ge 0} \pi \Big( \sigma b_m^2 \lambda_m   X_m 
    - b_m^1 \delta_m C_m \Big) \sigma^{2m}  \, , \label{Psi1} \\[5pt]
    \Psi_2 &= \sum_{m \ge 0}  b_m^1  \delta_m Z_m \sigma^{2m} \, , \label{Psi2}
\end{align}
\end{subequations}
where we have defined $\lambda_m = (2m+3)(2m+5) q_m^1 /32$ and $\delta_m = (m+1)(m+2) q_m^0 /8$, with $q_m^i = \Gamma \left( m+3+ i/2 \right)^{-2}$, $i = 1,2$.
In addition, $b_m^i = 2m+i$, $i = 1, \dots, 4$.
The series coefficients $X_m$, $C_m$, and $Z_m$ are functions of~$\mu^2$ only and are provided in Appendix~\ref{appendix:coeffs}.
Likewise, the corresponding series representation of the functions $\Psi_3$ and~$\Psi_4$ defining the kernel $\Gamma_2$ are 
\begin{subequations} \label{Psi3_Psi4}
    \begin{align}
	\Psi_3 &= \sum_{m\ge 0} 4\pi \Big( 
	b_m^3 \delta_m T_m 
	- \sigma b_m^4 \lambda_m  \, U_m \Big) \sigma^{2m+1} , \label{Psi3} \\[5pt]
	\Psi_4 &= \sum_{m\ge 0} 2 b_m^3  \delta_m G_m \sigma^{2m+1}  \, , \label{Psi4}
\end{align}
\end{subequations}
where $T_m$, $U_m$, and~$G_m$ are also functions of~$\mu^2$ only, the expressions of which are provided in Appendix~\ref{appendix:coeffs}.

\begin{figure}
    \centering
    \includegraphics[scale=0.52]{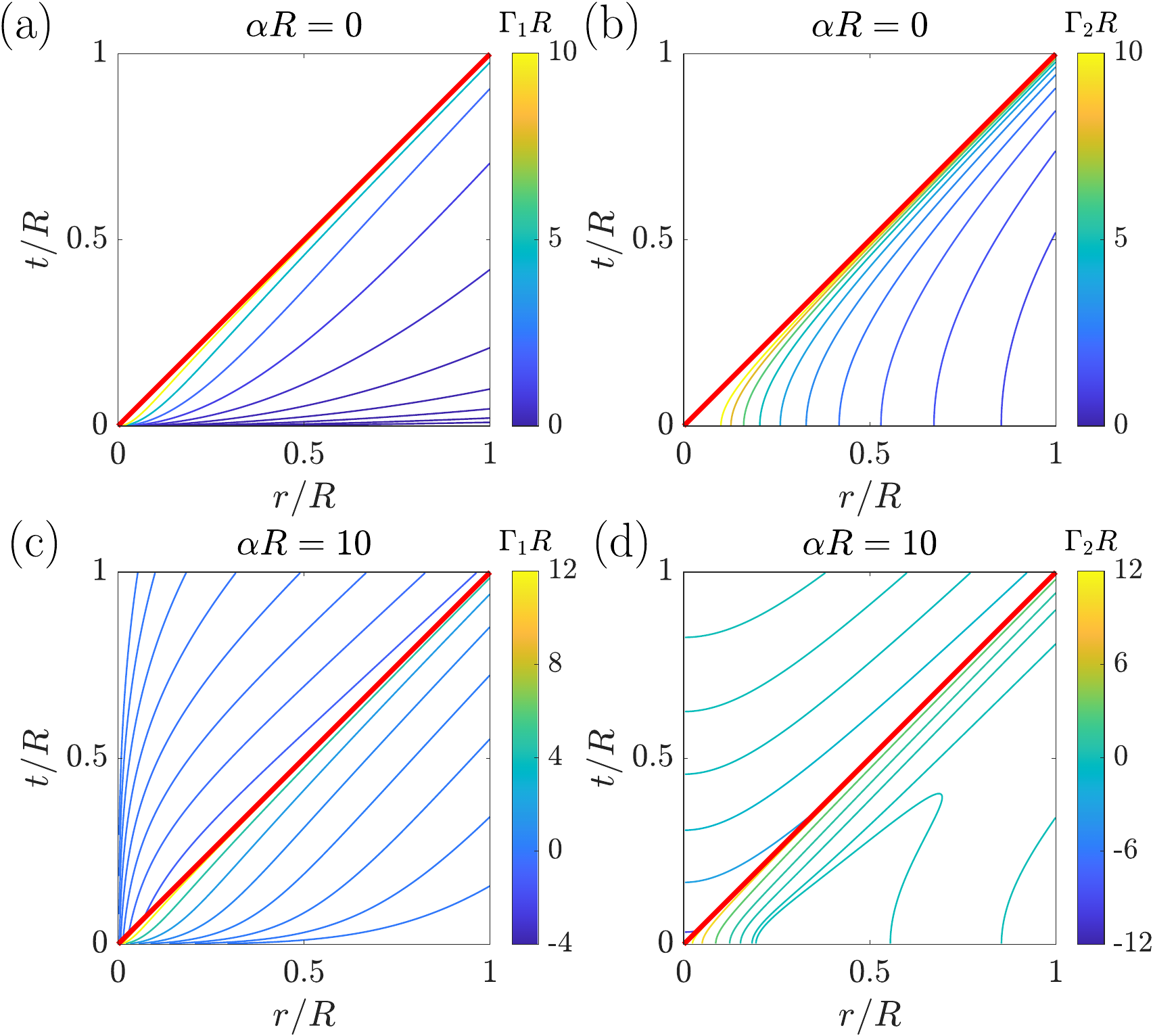}
    \caption{(Color online) Contour plots of the kernel functions (scaled by disk radius~$R$) for $\alpha R = 0$ [panels (a) and (b)] and $\alpha R = 10$ [panels (c) and (d)].
    Red solid line represent the first bisector corresponding to $t=r$.}
    \label{fig:contour}
\end{figure}

Equations~\eqref{Psi1_Psi2} and~\eqref{Psi3_Psi4} is a central result of our work.
Generally, only a few terms in the series expansion are needed to achieve an accurate estimation of the computed integrals for a given desired precision.
This approach is proven to be sometimes over a thousand times faster than the naive numerical evaluation of the infinite integrals computationally using standard techniques.  
A discussion concerning the truncation of the series expansion is provided in Appendix~\ref{appendix:trunc}.

In Fig.~\ref{fig:contour} we illustrate exemplary contour plots of the kernel functions~$\Gamma_1$ and~$\Gamma_2$ as defined by Eqs.~\eqref{Gammms}, rescaled by the radius of the disk~$R$.
Results are shown for two values of  $\alpha R = 0$ [panels (a) and (b)] and $\alpha R = 10$ [panels (c) and (d)].
Since $t,r \in [0,R]$, we present the arguments of the kernel functions in a scaled form by scaling $t$ and~$r$ by the relevant length scale corresponding to the radius of the disk.
Accordingly, the left-hand sides of the resulting Fredholm integral equations~\eqref{eq:dual-final} can conveniently be expressed as definite integrals on the interval $[0,1]$.
Notice that $t$ is a dummy variable used for integration.
The kernel functions $\Gamma_1$ and~$\Gamma_2$ have both dimensions of (length)$^{-1}$ thus these have been scaled likewise by~$R$ to be made dimensionless.
We note that $R$ and not~$h$ has been used here to scale lengths since the kernel functions are independent of~$h$.
Accordingly, the kernel functions have to be computed in a scaled form only once, not for each disk size, since their computation represents the most expensive part in solving the resulting integral equations.
The kernels are singular for $t=r$ (red lines), so that care must be taken when discretizing on a finer grid.
Unlike the Stokes flow limit for zero impermeability coefficient (or equivalently for an infinite permeability), the kernel functions for a Brinkman fluid take non-vanishing values for $r<t$ as well.

\subsection{Numerical solution of integral equations}

\begin{figure}
    \centering
    \includegraphics[scale=0.6]{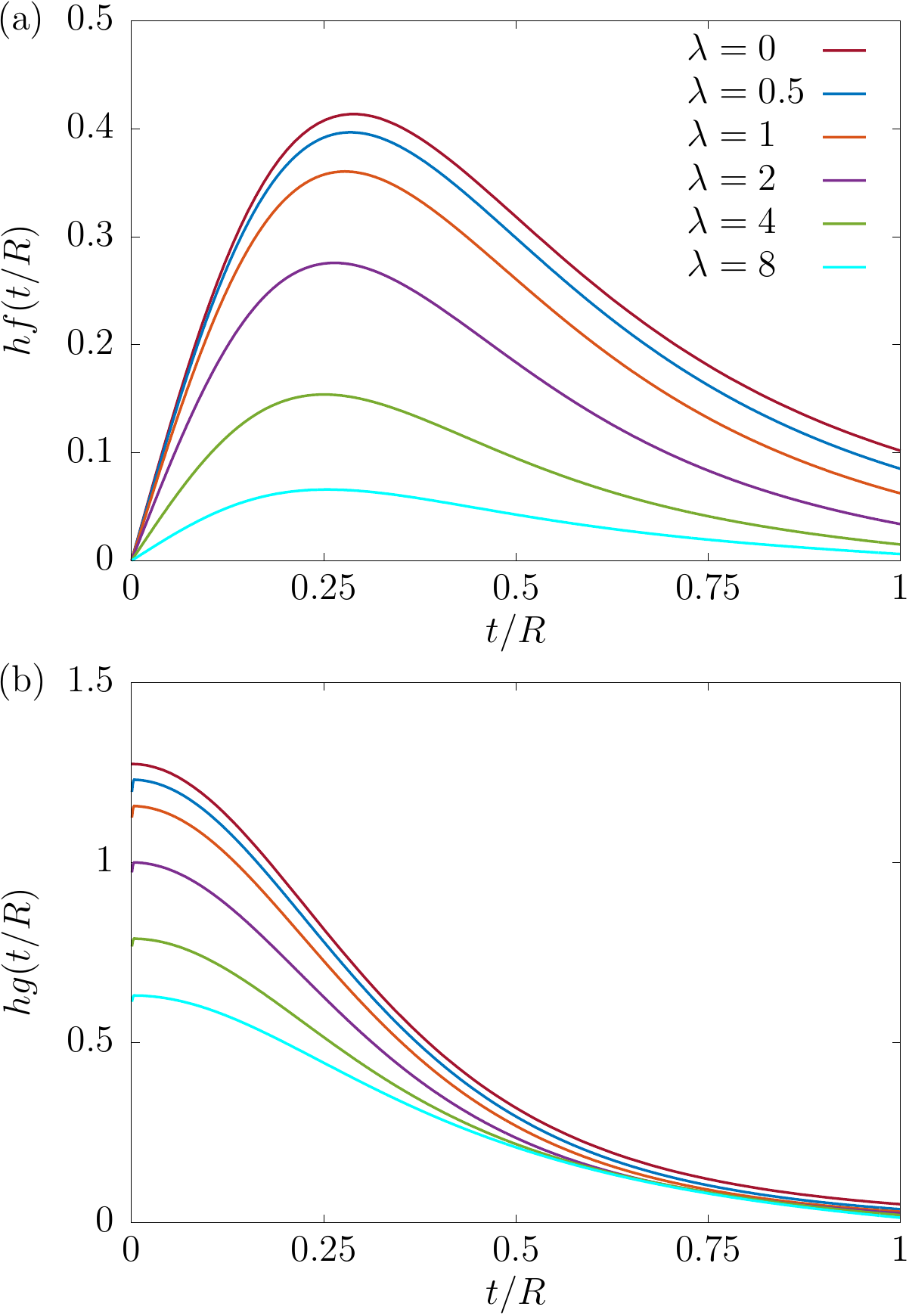}
    \caption{(Color online) Variation of the solutions of the Fredholm integral equations given by Eqs.~\eqref{eq:dual-final} as obtained numerically for various values of $\lambda = \alpha h$.
    Here, we set $\xi = h/R = 0.5$.}
    \label{fig:fg}
\end{figure}

In the following we summarize the main computational steps involved in the numerical solution of the resulting integral equations given by Eqs.~\eqref{eq:dual-final}.
The two integrals over the interval $[0, R]$ are partitioned into $N$~subintervals and approximated as midpoint Riemann sums.
Thereupon, the resulting equations are evaluated at $N$ collocation points $r_j$, $j = 1, \dots, N$, uniformly distributed over the interval $[0, R]$. 
Since for $t = r$ the kernel functions are singular, the partition of the sequence of points~$t_i$ used to approximate the integral as midpoint Riemann sums must be chosen differently from the collocation points~$r_j$.
Then, the discrete values of $f(t_i)$ and~$g(t_i)$ can readily be obtained by solving two well-behaved systems of $N$ linear equations numerically in Matlab~\cite{matlab22}.

We now define the dimensionless parameter $\lambda = \alpha h$ that we denominate as the porosity coefficient.
In addition, we define $\xi = h/R$ as the ratio between the singularity position and the radius of the disk.
Accordingly, the Stokes flow limit corresponds to $\lambda = 0$ while $\xi = 0$ holds in the limit of an infinitely extended no-slip wall.

In Fig.~\ref{fig:fg} we show the variation of the solutions of the Fredholm integral equations~\eqref{eq:dual-final} for various values of $\lambda$ while keeping $\xi = 0.5$.
Results are obtained numerically using $N=512$ discretization points.
In the limit $\lambda = 0$, we recover the solution previously obtained by Kim~\cite{kim83} in the Stokes flow limit.
Upon increasing the {porosity coefficient}, both functions amount to smaller values, implying that the magnitude of the velocity field becomes smaller as the flow resistance becomes larger. 
We can readily verify the fulfillment of the condition $f(0) = 0$ [Fig.~\ref{fig:fg}~(a)] for which the dual integral equation for the outer problem will be satisfied.

Once the solutions of the Fredholm integral equations for $f(t)$ and~$g(t)$ are obtained at $N$ discrete values, the velocity field can be obtained numerically at any point in the fluid domain.

\subsection{Expressions of the hydrodynamic fields}

By making use of the integral representation of the wavenumber-dependent coefficients $A(q)$ and~$B(q)$ given by Eqs.~\eqref{AB-solution-form}, the coefficients $c_1^\pm  = \left( B \pm A\right)/ \left( 2\alpha^2 \right)$ and $c_2^\pm = \left(\mp A-qB/Q\right)/ \left( 2\alpha^2 \right)$ can likewise be expressed in terms of definite integrals involving $f(t)$ and~$g(t)$.
By inserting the corresponding expressions of~$c_i^\pm$, $i=1,2$, into Eq.~\eqref{eq:image_final}, the image velocity and pressure fields can conveniently be presented in the form of definite integrals over the interval $[0, R]$ as
\begin{subequations} \label{eq:img_for_flow_field}
    \begin{align}
    G_r^\pm &= \int_0^R \big( \mathcal{K}_1 (r,z,t) f(t) \pm \mathcal{K}_2 (r,z,t) g(t) \big) \, \mathrm{d} t \, , \\
    G_z^\pm &= \int_0^R \big( \pm \mathcal{K}_3 (r,z,t) f(t) + \mathcal{K}_4 (r,z,t) g(t) \big) \, \mathrm{d} t \, , \label{eq:vzP_final} \\
    P^\pm &= \int_0^R \big( \mathcal{Q}_1 (r,z,t) f(t) \pm \mathcal{Q}_2 (r,z,t) g(t) \big) \, \mathrm{d} t \, ,
\end{align}
\end{subequations}
where $\mathcal{K}_i$, $i = 1, \dots, 4$, are given by 
\begin{subequations} \label{eq:Ks}
    \begin{align}
    \mathcal{K}_1 &= \frac{2}{\alpha^2} \int_0^\infty 
    q^2 \mathcal{S}_1 (q,z) \sin (qt) J_1(qr) \, \mathrm{d}q \, , \\
    \mathcal{K}_2 &= \frac{2}{\alpha^2} \int_0^\infty 
    q^2 \mathcal{S}_2 (q,z) \cos (qt) J_1(qr) \, \mathrm{d}q \, , \\
    \mathcal{K}_3 &= \frac{2}{\alpha^2} \int_0^\infty 
    q^2 \mathcal{S}_2 (q,z) \sin (qt) J_0(qr) \, \mathrm{d}q \, , \\
    \mathcal{K}_4 &= \frac{2}{\alpha^2} \int_0^\infty 
    q^2 \mathcal{S}_3 (q,z) \cos (qt) J_0(qr) \, \mathrm{d}q \, .
\end{align}
\end{subequations}
In addition, $\mathcal{Q}_1$ and $\mathcal{Q}_2$ are both independent of~$\alpha$ and are given by
\begin{subequations}  \label{eq:Qs}
    \begin{align}
    \mathcal{Q}_1 &= - \int_0^\infty
    q e^{-q|z|} \sin(qt) J_0(qr) \, \mathrm{d} q \, , \\
    \mathcal{Q}_2 &= - \int_0^\infty
    q e^{-q|z|} \cos(qt) J_0(qr) \, \mathrm{d} q \, .
\end{align}
\end{subequations}
Here, we have defined $\mathcal{S}_1 = \left( Q/q\right) e^{-Q|z|} - e^{-q|z|}$, $\mathcal{S}_2 = e^{-Q|z|} - e^{-q|z|}$, and $\mathcal{S}_3 = \left( q/Q\right) e^{-Q|z|} - e^{-q|z|}$.
The corresponding expressions of $\mathcal{K}_i$, $i=1, \dots, 4$, in the limit $\lambda \to 0$ are provided in Appendix~\ref{appendix:exp}.

For an efficient numerical computation, the six infinite integrals stated by Eqs.~\eqref{eq:Ks} and~\eqref{eq:Qs} are converted into well-behaved definite integrals over the domain $[0, \pi]$ by using the variable substitution $q = \tan \left( \frac{u}{2} \right)$ and thus $\mathrm{d}q = \frac{1}{2} \left( 1+q^2\right) \mathrm{d}u$.
Thereupon, the resulting definite integrals are approximated as midpoint Riemann sums.
Finally, the image solution is computed via Eqs.~\eqref{eq:img_for_flow_field} again by approximating the definite integral as midpoint Riemann sums in the interval $[0,R]$.
For an improved numerical evaluation of the flow field, we typically discretize the interval $[0, \pi]$ into $M = 100N$ points uniformly distributed over the definite interval.

\begin{figure}
    \centering
    \includegraphics[scale=0.36]{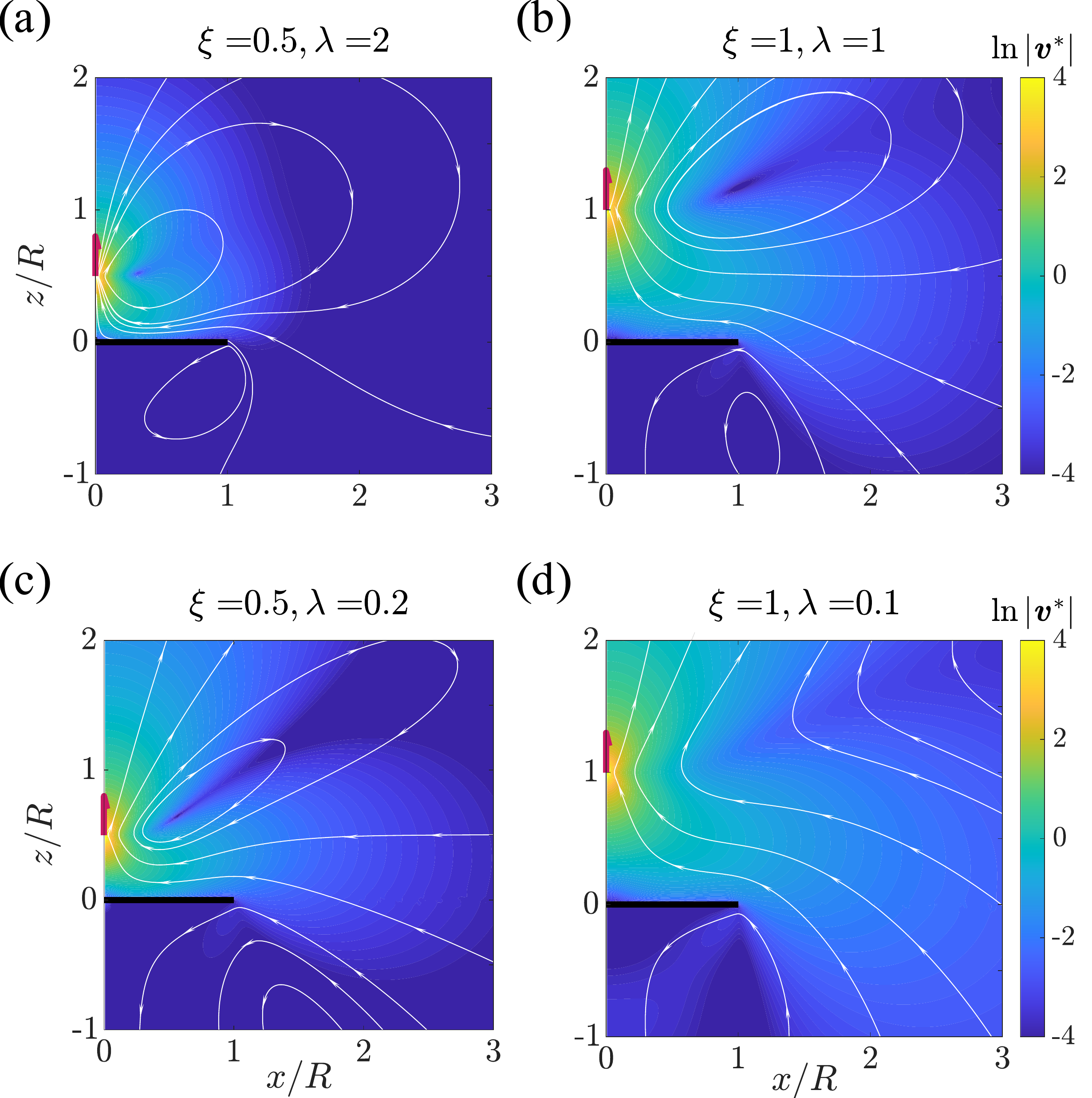}
    \caption{(Color online) Streamlines and contour plots of the scaled velocity field induced by Brinkmanlet for $\xi = 0.5$ [(a) and~(c)], and $\xi = 1$ [(b) and~(d)].
    Here, $\vect{v}^* = \left. \vect{v} \middle/ \left( F/ \left( 8\pi\eta R \right) \right) \right.$ denotes the dimensionless velocity.
    }
    \label{fig:monopole}
\end{figure}

The flow streamlines and contour plots of the velocity magnitude are shown in Fig.~\ref{fig:monopole} for various values of $\xi$ and~$\lambda$.
For larger values of the porosity coefficient~$\lambda$, we observe the formation of eddies with closed streamlines in the fluid region above the plate [panels~(a), (b), and~(c)].
For the present set of parameters, these eddies do not occur for smaller $\lambda$ [panel~(d)] or in the Stokes flow limit as $\lambda \to 0$ [c.f.~Ref.~\onlinecite{kim83}].
Accordingly, the flow behavior in a Brinkman fluid qualitatively differs from that observed for a Stokes flow.
Upon increasing~$\lambda$, the resistance of the flow becomes larger, so that the magnitude of the flow velocity field undergoes a sharp decay with distance from the singularity position.

\subsection{Solutions in the liming cases}

To connect our work with previous studies, we recall the following known solutions obtained in the limiting cases of an infinitely extended plate $\left(R\to\infty\right)$ or zero impermeability $(\lambda \to 0)$.

\paragraph{Solution for an infinitely extended plate}

In the following, we recover the solution of the flow problem in the limiting case of an infinitely extended plate such that $R\to\infty$.
In this situation, the solution can be obtained via inverse Hankel transform of Eqs.~\eqref{inner-problem} to obtain 
\begin{subequations}
    \begin{align}
    A(q) &= -\frac{2 q \alpha^2}{Q-q} \int_0^\infty G_r^\infty (r,z=0) \, J_1 (qr) \, r \, \mathrm{d} r \, , \\
    B(q) &= -\frac{2 Q \alpha^2}{Q-q} \int_0^\infty G_z^\infty (r,z=0) \, J_0 (qr) \, r \, \mathrm{d} r \, .
\end{align}
\end{subequations}
The latter two integrals can be evalted analytically as
\begin{subequations} \label{eq:A_B_infR}
    \begin{align}
    A(q) &= -\frac{4 q^2}{Q-q} \left( e^{-qh} - e^{-Qh} \right) \, , \\
    B(q) &= -\frac{4 q}{Q-q} \left( Q e^{-qh} - q e^{-Qh} \right) \, .
    \end{align}
\end{subequations}
Then, the wavenumber-dependent coefficients $c_i^\pm$, $i = 1,2$, are obtained as
\begin{subequations}
    \begin{align}
    c_1^+  &= \frac{2q}{\alpha^2 \left( Q-q \right)} \left( 2q e^{-Qh} - \left( Q+q\right)e^{-qh} \right) \, , \\
    c_2^+   &= \frac{2q^2}{\alpha^2 Q\left( Q-q \right)} \left( 2Q e^{-qh} - \left( Q+q\right)e^{-Qh} \right) \, .
\end{align}
\end{subequations}
These results are in full agreement with those obtained by Felderhof~\cite{felderhof05, felderhof2009flow} and later rederived in Ref.~\onlinecite{fouxon2018fundamental} using to some degree a different solution approach.
Particularly, in the limit $\alpha \to 0$, it follows that $Q \sim q + \alpha^2/ \left( 2q\right)$.
We readily obtain $A(q) = -4q^2 h e^{-qh}$ and $B(q) = -4q \left( 1+qh\right) e^{-qh}$, corresponding to the solution originally obtained by Blake~\cite{blake71}.

We note that in the limit $R \to \infty$, the expressions of $f(t)$ and~$g(t)$ can be obtained from Eqs.~\eqref{AB-solution-form} via inverse sine and cosine Fourier transform as
\begin{subequations} \label{fg}
    \begin{align}
    f(t) &= -\frac{1}{2\pi} \int_0^\infty \frac{A(q)}{q} \, \sin (qt) \, \mathrm{d} q \, , \\
    g(t) &= -\frac{1}{2\pi} \int_0^\infty \frac{B(q)}{q} \, \cos (qt) \, \mathrm{d} q \, .
\end{align}
\end{subequations}
Then, inserting the expressions of $A(q)$ and~$B(q)$ given by Eqs.~\eqref{eq:A_B_infR} into Eqs.~\eqref{fg} leads to
\begin{equation} \label{eq:fg}
f(t) = \frac{4}{\pi} \frac{h^2 t}{\left( t^2+h^2 \right)^2} \, , \quad 
g(t) = \frac{4}{\pi} \frac{h^3}{\left( t^2+h^2 \right)^2} \, .
\end{equation}

\paragraph{Solution in the Stokes limit}

In the limit $\lambda \to 0$, Fredholm integral equations~\eqref{eq:dual-final} reduce to classical Abel integral equations.
These represent particular cases of Volterra integral equations.
We thereby recover the same integral equations obtained previously by Kim~\cite{kim83} who examined the axisymmetric Stokes flow near a finite-sized disk using a dual integral equation approach. 
Specifically,
\begin{subequations} \label{eq:kim-dual}
\begin{align}
	\int_0^r \frac{t f(t) \, \mathrm{d}t}{r\left( r^2-t^2 \right)^\frac{1}{2}} &= \frac{hr}{\left(r^2 + h^2 \right)^\frac{3}{2}} \, , \\ 
	\int_0^r \frac{g(t) \, \mathrm{d}t}{\left( r^2-t^2 \right)^\frac{1}{2}} &= \frac{ r^2 + 2 h^2 }{\left(r^2 + h^2 \right)^\frac{3}{2}} \, .
\end{align}
\end{subequations}

Surprisingly, integrating Eqs.~\eqref{eq:kim-dual} yields exactly the same expressions of $f(t)$ and~$g(t)$ previously obtained for an infinitely extended plate given by Eq.~\eqref{eq:fg}.
We have checked that this peculiar property only hold for $\alpha = 0$ and does not hold for a Brinkman fluid, for which the solution should be obtained by systematically solving the resulting Fredholm dual integral equations numerically.

\subsection{Hydrodynamic monopole reaction}

Having presented a semi-analytical theory describing the axisymmetric monopole flow of Brinkman fluid near a stationary plate of circular shape, we next assess the effect of confining plate on the slow motion of a colloidal particle moving along the symmetry axis of the plate. 
We define the scaled normal-normal component of the reaction tensor as
\begin{equation}
    \mathcal{R} = \frac{3}{4} \, h \,\lim_{(r,z) \to (0,h)} G_z^+ \, , 
\end{equation}
where the factor $3/4$ results from the fact that the Green's function is usually scaled by the bulk hydrodynamic mobility coefficient $6\pi\eta a$.
It follows from Eq.~\eqref{eq:vzP_final} that
\begin{equation}
    \mathcal{R} = \frac{3}{4} \, h \int_0^R 
    \bigg( \mathcal{K}_3(0,h,t) f(t) + \mathcal{K}_4 (0,h,t) g(t) \bigg) \, \mathrm{d} t \, . \label{eq:raction}
\end{equation}
The latter integral can be approximated and evaluated numerically using a standard midpoint Riemann sum.

We will provide the following exact analytical expressions in the limiting cases of $R\to\infty$ and~$\lambda \to 0$.

\paragraph{Solution for an infinitely extended plate}

In the limit $R \to \infty$, Eq.~\eqref{eq:raction} becomes an infinite integral that can be evaluated analytically and cast in the form
\begin{align}
        \mathcal{R} &= - 
         \big( \Lambda_1 + \Lambda_2 e^{-2\lambda} - \Lambda_3 e^{-\lambda} + \Lambda_4 E_1(\lambda) + \Lambda_5 K_0(2\lambda) \notag \\
        &\quad \left.  + \, \Lambda_6 K_1(2\lambda) + \Lambda_7 N_0(2\lambda) - \Lambda_8 N_1(2\lambda) \big) \middle/ \left( 16\lambda^4 \right) \right. , \label{federeaktion}
   \end{align}
where we have defined the abbreviation $N_\nu(z) = Y_\nu(z) - H_\nu(z)$.
Here, $Y_\nu$ and~$K_\nu$ denote the Bessel function of the second kind and modified Bessel function of the second kind, respectively, $H_\nu$ is the Struve function, and $E_1$ is related to the exponential integral $\operatorname{Ei}$ for a real argument via $E_1(\lambda) = -\operatorname{Ei}(-\lambda)$.
In addition, $\Lambda_i$, $i = 1, \dots, 8$, are functions of~$\lambda$ only.
They are given by $\Lambda_1 = 6 \left( 6+\lambda^2+8\lambda^3 \right)$, $\Lambda_2 = 6 \left( 6+12\lambda+9\lambda^2+2\lambda^3 \right)$, $\Lambda_3 = 144+144\lambda+48\lambda^2-6\lambda^4+10\lambda^5+\lambda^6-\lambda^7$, $\Lambda_4 = \lambda^6 \left( 12-\lambda^2\right)$, $\Lambda_5 = 72 \lambda^2$, $\Lambda_6 = 36\lambda(2+\lambda^2)$, $\Lambda_7 = 12\pi \lambda^2 \left( 3-\lambda^2 \right)$, and $\Lambda_8 = 6\pi \lambda \left( 6-5\lambda^2 \right)$.
We note that Ref.~\onlinecite{felderhof05jcp} reports some of the coefficients~$\Lambda_i$ with erroneous prefactors that we correct here.

For $\lambda \ll 1$, the reaction can be expanded in power series of~$\lambda$ as
\begin{equation}
    \mathcal{R} = -\frac{9}{8} + \lambda - \frac{3}{8} \, \lambda^2 + \frac{19}{192} \, \lambda^4 - \frac{1}{10}\, \lambda^5 + \mathcal{O} \left( \lambda^6 \right), \notag
\end{equation}
which is identical to the result obtained by Felderhof~\cite[Eq.~(3.12)]{felderhof05jcp}.
We also recognize the leading-order correction to the reaction tensor for motion perpendicular to a planar wall as first obtained by Lorentz~\cite{lorentz07}.
In the limit $\lambda \gg 1$, a series expansion in inverse powers of~$\lambda$ yields
\begin{equation}
    \mathcal{R} = -\frac{3}{8} \, \lambda^{-2}-\frac{9}{8} \, \lambda^{-3}-\frac{9}{4} \, \lambda^{-4}-\frac{45}{16} \, \lambda^{-5} + \mathcal{O} \left(\lambda^{-7} \right) \, . \notag 
\end{equation}

\begin{figure}
    \centering
\includegraphics[scale=0.7]{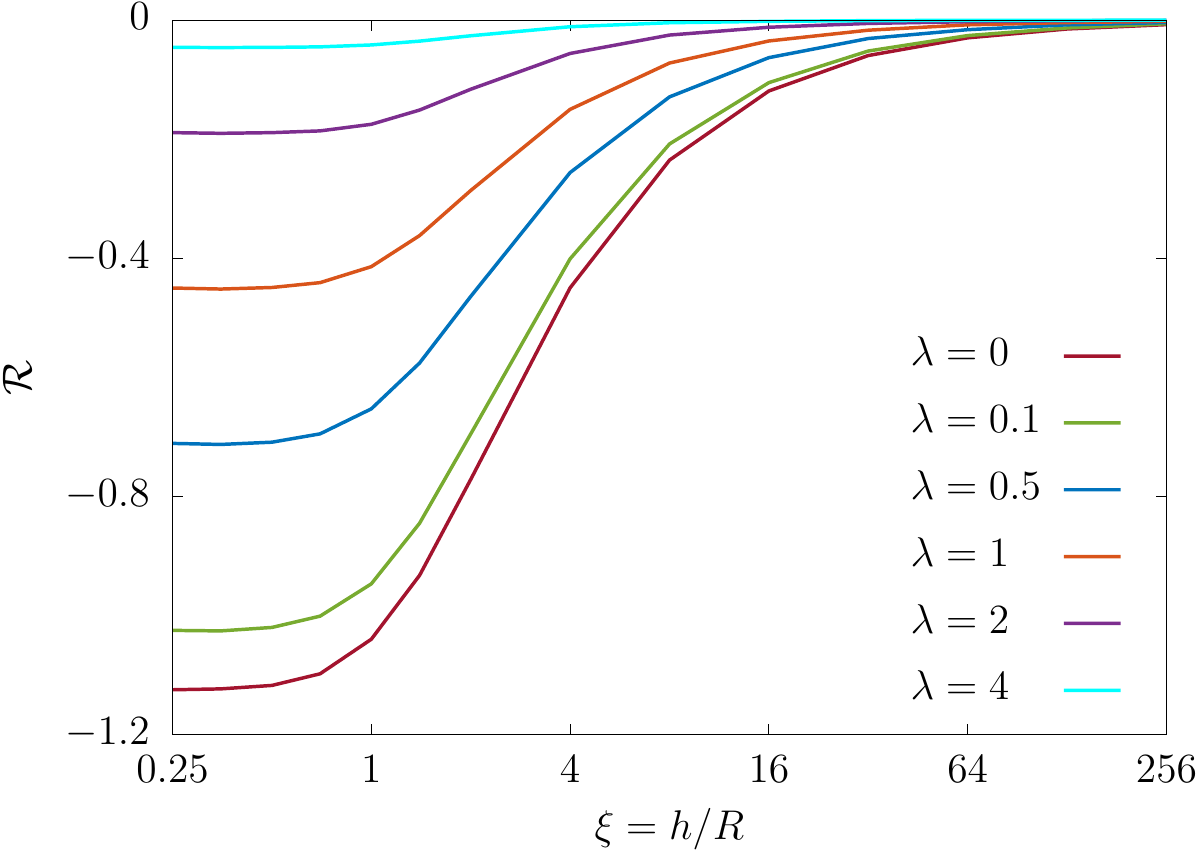}
    \caption{(Color online) Variation of the hydrodynamic monopole reaction versus the ratio~$\xi = h/R$ for various values of the porosity coefficient~$\lambda = \alpha h$ as obtained by numerically integrating Eq.~\eqref{eq:raction}.
    The red curve representing $\lambda=0$ corresponds to the Stokes flow limit given in a closed form by Eq.~\eqref{kimreaktion}.
    }
    \label{fig:reaction}
\end{figure}

\paragraph{Solution in the Stokes limit}

In the limit $\lambda \to 0$, the reaction defined by Eq.~\eqref{eq:raction} is expressed as a double integral of the form
\begin{equation}
    \mathcal{R} = -\frac{3}{\pi} \, h^4 \int_0^\infty e^{-qh} \, \mathrm{d} q
    \int_0^R \tau (q, t)  \, \mathrm{d} t \, , 
\end{equation}
where
\begin{equation}
    \tau (q, t) = \frac{qt \sin(qt)+\left(1+qh\right)\cos(qt)}{\left( t^2+h^2\right)^2} \, .
\end{equation}

By swapping the order of integration, and performing first integration with respect to~$q$, and then with respect to~$t$, the reaction takes the final compact form
\begin{equation}
    \mathcal{R} = -\frac{3}{4\pi} 
    \left( \frac{\xi \left( 3+5\xi^2 \right)}{\left( 1+\xi^2\right)^2} + 3 \arctan \left( \xi^{-1} \right) \right) , \label{kimreaktion}
\end{equation}
where again $\xi = h/R$.
The result is in full agreement with the expression given by Kim~\cite{kim83}.
In the limit $\xi \ll 1$, 
\begin{equation}
    \mathcal{R} = -\frac{9}{8} + \frac{6}{5\pi} \, \xi^5 - \frac{18}{7\pi} \, \xi^7 + \frac{4}{\pi} \, \xi^9 
    + \mathcal{O} \left( \xi^{11} \right) . \notag
\end{equation}
In the limit $\xi \gg 1$, we obtain
\begin{equation}
    \mathcal{R} = -\frac{6}{\pi} \, \xi^{-1} + \frac{6}{\pi} \, \xi^{-3} - \frac{36}{5\pi} \, \xi^{-5} + \frac{60}{7\pi} \, \xi^{-7}
    + \mathcal{O} \left( \xi^{-9} \right) . \notag 
\end{equation}

In Fig.~\ref{fig:reaction} we present the variation of the monopole reaction as a function of~$\xi$ for various values of~$\lambda$.
The magnitude of the reaction varies monotonically upon varying the system size, reaching a maximum value given by Eq.~\eqref{federeaktion} in the limit $\xi \to 0$ corresponding to an infinitely extended plate. 
Upon increasing the porosity coefficient the magnitude of the reaction monotonically decreases.

\section{Dipolar flow field}

Having derived the solution for the axisymmetric monopole flow induced by Brinkmanlet located on the symmetry axis of a stationary no-slip disk, we now make use of this fundamental solution to determine the corresponding axisymmetric dipolar flow field.
The latter is obtained by taking the derivative of the monopole flow field with respect to the singularity position~\cite{lopez2014dynamics, sprenger2020towards}.
For an axisymmetric configuration, the dipole flow field can obtained as
\begin{equation}
    \vect{v}_\mathrm{D} = \beta \left( \vect{G}_\mathrm{D}^\infty + \vect{G}_\mathrm{D} \right) \, , 
\end{equation}
wherein 
\begin{equation}
    \vect{G}_\mathrm{D}^\infty = \frac{\partial \vect{G}^\infty}{\partial h} \, , \qquad
    \vect{G}_\mathrm{D} = \frac{\partial \vect{G}}{\partial h}
\end{equation}
stand for the free-space- and image-dipole-related contributions, respectively.
Here,~$\beta$ denotes the dipole coefficient which has dimension of (length)$^3$(time)$^{-1}$.

The viscous flow field induced by a dipole is the leading contribution of many self-propelling active microswimmers, which are by definition force free in the inertialess regime of swimming. 
To achieve self propulsion, a large variety of bacterial microorganisms, such as \textit{E. coli}, leverage bundles of helical filaments known as flagella, whose rotation causes the entire bacterium body to move forward in a corkscrew-like motion.
Since this type of microswimmers push out fluid along their swimming axis, they are known as pushers. 
Another broad class of swimmers referred to as pullers, such as the single-cell green alga \textit{Chlamydomonas reinhardtii}, pull in fluid along their swimming direction and repel fluid from the sides of their bodies.
Depending on the sign of the dipolar coefficient~$\beta$, we discriminate between pushers $\left( \beta > 0 \right)$ and pullers $\left( \beta < 0 \right)$.
Depending on the shape profile of microswimmers, it has been shown that the optimal swimmer can be a puller, pusher or neutral~\cite{daddi2021optimal}.
In the presence of confining interfaces, the locomotory behavior and swimming trajectories are known to depend on the swimmer type in a complex fashion~\cite{ishimoto2013squirmer, li2014hydrodynamic, daddi2018state, daddi2018swimming, daddi2020tuning, daddi2021hydrodynamics}.

\subsection{Dual integral equations}

Since an analytical solution of the integral equations for $f(t)$ and~$g(t)$ was shown to be delicate, computing the derivative with respect to~$h$ is rather not trivial in any event.
For the derivation of the dipole flow field, we follow an analytical framework analogous to that employed for the determination of the monopole flow field.
Specifically, we write the solution of the flow problem as a superposition of the free-space dipole $\vect{G}_\mathrm{D}^\infty$ and a complementary solution $\vect{G}_\mathrm{D}$ that is required to satisfy the no-slip boundary condition on the disk.
The problem can likewise be formulated as a usual mixed boundary value problem that is also transformed into dual integral equations.
Accordingly, the problem reduces to expressing the unknown wavenumber-dependent coefficients $A_\mathrm{D} (q)$ and~$B_\mathrm{D} (q)$ in terms of definite integrals of the form
\begin{subequations} \label{AB-solution-form-dipole}
\begin{align}
	A_\mathrm{D}(q) &= -4 q \int_0^R f_\mathrm{D}(t) \sin (qt) \, \mathrm{d} t \, , \\
	B_\mathrm{D}(q) &= -4 q\int_0^R g_\mathrm{D}(t) \cos (qt) \, \mathrm{d} t \, ,
\end{align}
\end{subequations}
where $f_\mathrm{D}(t)$ and~$g_\mathrm{D}(t)$, $t \in [0, R]$ needs to be determined by solving the integral equations for the inner problem
\begin{subequations}
\begin{align}
	\int_0^R f_\mathrm{D}(t) \Gamma_1 (r,t) \, \mathrm{d}t &= F_1(r)  \, , \\[2pt]
	\int_0^R g_\mathrm{D}(t) \Gamma_2 (r,t) \, \mathrm{d}t &= F_2(r) \, ,
\end{align}
\end{subequations}
with the known radial functions on the right-hand side 
\begin{align}
    F_1 (r) &= \frac{r}{\rho^3} \left( \left(1-3 \left( \frac{h}{\rho}\right)^2 \right) \beta_2
    + h \, \frac{\partial \beta_2}{\partial h}
    \right) , \notag \\
    F_2(r) &= \frac{h}{\rho^3} \left( \left( 2-3 \left( \frac{h}{\rho} \right)^2 \right) \beta_2 + h\, \frac{\partial \beta_2}{\partial h} - \beta_1 \right)+
    \frac{1}{\rho} \frac{\partial \beta_1}{\partial h} . \notag
\end{align}
Here, we have used the fact that $\partial \rho / \partial h = h/\rho$.
Mathematically, $f_\mathrm{D} (t) = \partial f (t) / \partial h$ and $g_\mathrm{D} (t) = \partial g (t) / \partial h$.
In the limits $\alpha \to 0$ or $R \to \infty$, we obtain 
\begin{equation}
    f_\mathrm{D}(t) = \frac{8}{\pi}
    \frac{ht \left( t^2-h^2 \right)}{\left( t^2+h^2 \right)^3} \, , \quad
    g_\mathrm{D}(t) = \frac{4}{\pi} 
    \frac{h^2 \left( 3t^2-h^2 \right)}{\left( t^2+h^2 \right)^3} \, .
    \label{eq:f_D_g_D}
\end{equation}

\begin{figure}
    \centering
    \includegraphics[scale=0.36]{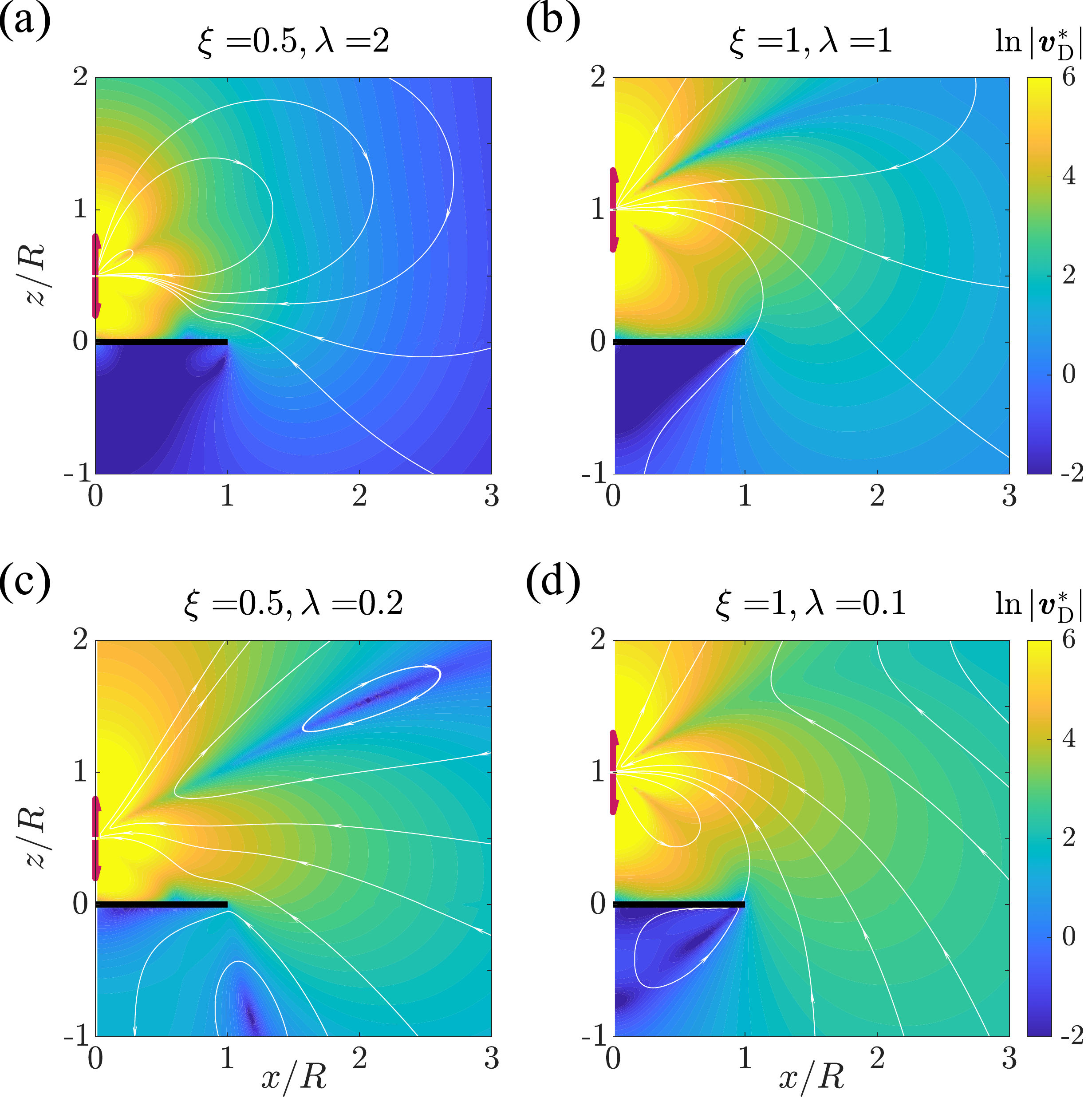}
    \caption{(Color online) Streamlines and contour plots of the scaled velocity field induced by dipole singularity for $\xi = 0.5$ [(a) and~(c)], and $\xi = 1$ [(b) and~(d)].
    Here, $\vect{v}_\mathrm{D}^* = \left. \vect{v}_\mathrm{D} \middle/ \left( \beta/R^2 \right) \right.$.
    }
    \label{fig:dipole}
\end{figure}

Figure~\ref{fig:dipole} shows the resulting flow streamlines and contour plots of the dipolar velocity magnitude for various values of~$\xi$ and~$\lambda$.
The flow structure and eddy formation depend strongly on the underlying physical parameters. 
Likewise, the magnitude of the self-induced dipolar flow decreases upon increasing the porosity coefficient, leading to a screened flow field far away from the singularity position.

\subsection{Hydrodynamic dipole reaction}

We next assess the effect of the confining disk on the axisymmetric motion of a microswimmer through the calculation of the normal-normal component of the reaction tensor that we define as
\begin{equation}
    \mathcal{R}_\mathrm{D} = \frac{3}{4} \, h^2 \, \lim_{(r,z) \to (0,h)}
    {G_\mathrm{D}}^+_z \, .
\end{equation}
The latter can be cast in the integral form
\begin{equation}
    \mathcal{R}_\mathrm{D} = \frac{3}{4} \, h^2 \int_0^R 
    \bigg( \mathcal{K}_3(0,h,t) f_\mathrm{D}(t) + \mathcal{K}_4 (0,h,t) g_\mathrm{D}(t) \bigg) \mathrm{d} t , \label{eq:raction_DIPOLE}
\end{equation}
which can likewise be approximated and evaluated numerically using a standard midpoint Riemann sum.

We now provide the corresponding expressions in the limiting cases of $\xi \to 0$ and $\lambda \to 0$.

\paragraph{Solution for an infinitely extended plate}

In the limit $R \to \infty$, the resulting infinite integral given by Eq.~\eqref{eq:raction_DIPOLE} can be evaluated analytically as
\begin{align}
	\mathcal{R}_\mathrm{D} &= 
	\big( \Pi_1 + \Pi_2 e^{-2\lambda} -\Pi_3 e^{-\lambda} + \Pi_4 E_1(\lambda) + \Pi_5 K_0(2\lambda) \notag \\
	&\, \left. +\,  \Pi_6 K_1 (2\lambda) + \Pi_7 N_0(2\lambda) - \Pi_8 N_1(2\lambda) \big) \middle/ \left( 64 \lambda^4 \right) \right. ,
	\label{dipolereaktionxi}
\end{align}
wherein $\Pi_i$, $i = 1, \dots, 8$, are functions of~$\lambda$ and are explicitly given by $\Pi_1 = 12 \left( 30+3\lambda^2+40\lambda^3-8\lambda^5 \right)$, $\Pi_2 = 12 \left( 30+60\lambda+51\lambda^2+22\lambda^3+4\lambda^4 \right)$, $\Pi_3 = 6 \left( 240+240\lambda+96\lambda^2+16\lambda^3-2\lambda^4-2\lambda^5-\lambda^6+\lambda^7 \right)$, $\Pi_4 = 6\lambda^6 \left( \lambda^2-4 \right)$, $\Pi_5 = 144 \lambda^2 \left( 5+\lambda^2\right)$, $\Pi_6 = 72 \lambda \left( 10 + 7\lambda^2 \right)$, $\Pi_7 = 72\pi \lambda^2 \left( 5-2\lambda^2 \right)$, and $\Pi_8 = 12\pi \lambda \left( 30-27\lambda^2+4\lambda^4 \right)$.
We recall the abbreviation $N_\nu(z) = Y_\nu(z) - H_\nu(z)$.
We believe that the closed form expression given by Eq.~\eqref{dipolereaktionxi} is original and has not been reported in the literature so far.

In the limit $\lambda \ll 1$, we obtain
\begin{equation}
    \mathcal{R}_\mathrm{D} = \frac{9}{16} - \frac{3}{16} \, \lambda^2 + \frac{19}{128} \, \lambda^4
    - \frac{1}{5} \, \lambda^5 + \mathcal{O} \left( \lambda^6 \right) \, . \notag 
\end{equation}
We also recognize the leading-order contribution to the induced swimming speed due to dipolar hydrodynamic interactions with a plane surface~\cite{berke2008hydrodynamic, spagnolie2012hydrodynamics, mathijssen2015hydrodynamics, daddi19pre}.

For $\lambda \gg 1$, we obtain
\begin{equation}
    \mathcal{R}_\mathrm{D} = \frac{9}{16} \, \lambda^{-2} + \frac{9}{4} \, \lambda^{-3} + \frac{45}{8} \, \lambda^{-4} + \frac{135}{16} \, \lambda^{-5}
    + \mathcal{O} \left( \lambda^{-7} \right) . \notag 
\end{equation}

\begin{figure}
    \centering
\includegraphics[scale=0.7]{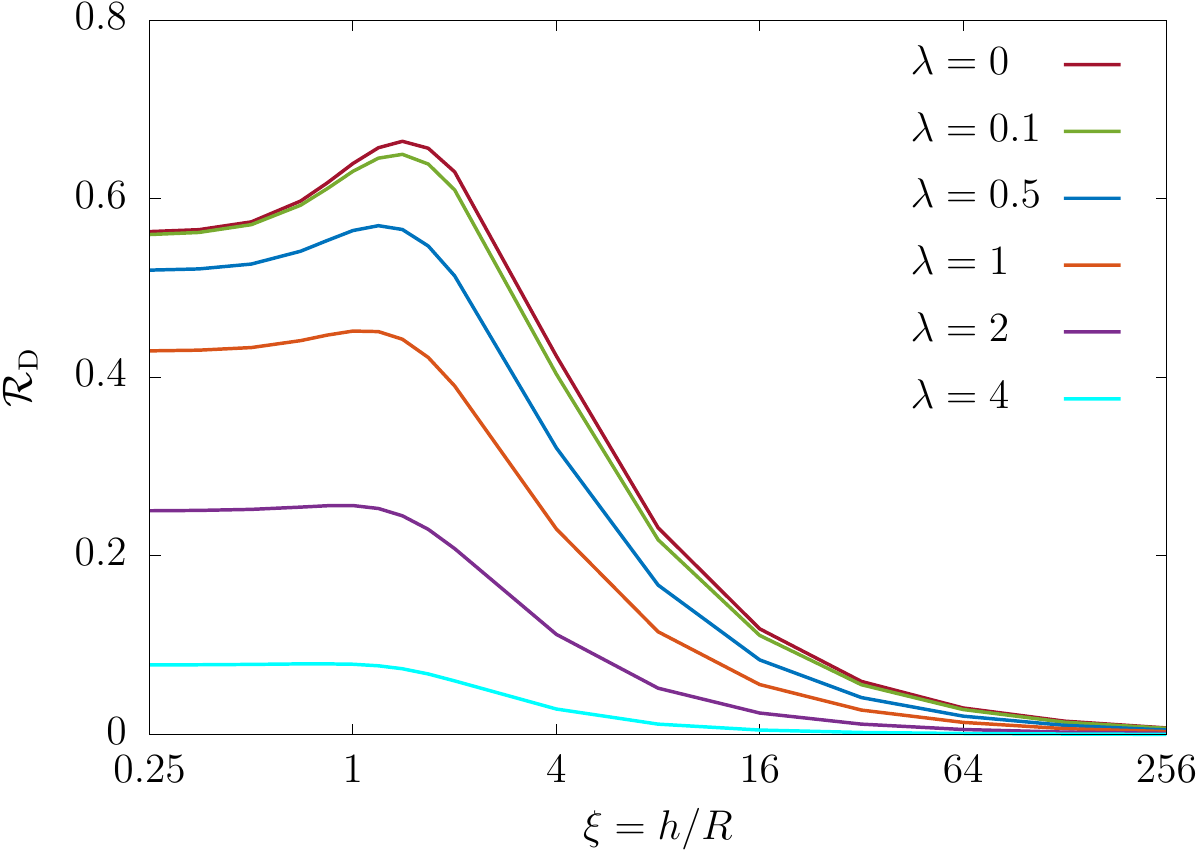}
    \caption{(Color online) Variation of the hydrodynamic dipole reaction versus $\xi = h/R$ for various values of the porosity coefficient~$\lambda = \alpha h$ as obtained by numerically integrating Eq.~\eqref{eq:raction_DIPOLE}.
    The reaction for $\lambda=0$ (red) corresponds to the Stokes limit given in a closed form by Eq.~\eqref{dipolereaktionlambda}.
    }
    \label{fig:reaction-dipole}
\end{figure}

\paragraph{Solution in the Stokes limit}

In the limit $\lambda \to 0$, an exact analytical expression of the dipole reaction can be obtained.
By inserting the expressions of $f_\mathrm{D}(t)$ and $g_\mathrm{D}(t)$ stated by Eq.~\eqref{eq:f_D_g_D} into Eq.~\eqref{eq:raction_DIPOLE} and performing the integration with respect to~$q$ and then with respect to~$t$, we obtain 
\begin{equation}
    \mathcal{R}_\mathrm{D} = \frac{3}{8\pi} 
    \left( \frac{\xi \left( 3+8\xi^2+13\xi^4 \right)}{\left(1+\xi^2 \right)^3} + 3 \arctan \left( \xi^{-1}\right) \right) .
    \label{dipolereaktionlambda}
\end{equation}

Particularly, for $\xi \ll 1$, an expansion in power series of~$\lambda$ gives
\begin{equation}
    \mathcal{R} = \frac{9}{16} + \frac{12}{5\pi} \, \xi^5 - \frac{54}{7\pi} \, \xi^7 + \frac{16}{\pi} \, \xi^9
    + \mathcal{O} \left( \xi^{11} \right) . \notag 
\end{equation}

For $\xi \gg 1$, we obtain
\begin{equation}
    \mathcal{R} = \frac{6}{\pi} \, \xi^{-1} - \frac{12}{\pi} \, \xi^{-3} + \frac{108}{5\pi} \, \xi^{-5} - \frac{240}{7\pi} \, \xi^{-7}
    + \mathcal{O} \left( \xi^{-9} \right). \notag 
\end{equation}

In Fig.~\ref{fig:reaction-dipole} we present the hydrodynamic dipole reaction versus $\xi$ for different values of~$\lambda$.
Results are obtained by integrating Eq.~\eqref{eq:raction_DIPOLE} numerically. 
Unlike the monopole reaction which has been shown to follow a monotonic behavior upon varying the system size, the dipole reaction displays an peak value around $\xi \sim 1$.
Accordingly, the effect of the confining disk on the locomotory behavior of a self-propelling dipole swimmer is expected to be larger around this value.
A qualitative understanding of the observed peak can be gained from the following consideration. In a Stokes fluid, a force dipole induces a flow field that changes direction at an angle $\theta_m=\arctan \sqrt{2}$ with respect to the symmetry axis, also known as the ``magic angle''. One can expect that the reaction due to the disk becomes stronger with increasing disk radius $R$, but only as long as the disk does not reach across the flow reversal cone. For larger radii, the reaction due to the disk decreases again. 
In a Brinkman fluid, the flow reversal occurs at a larger angle and the peak in the reaction moves towards larger radii (smaller $\xi$) and diminishes.

\section{Discussion and concluding remarks}

In the present manuscript, we have presented a semi-analytical theory describing the axisymmetric low-Reynolds-number flows induced by monopole and dipole singularities in proximity of a no-slip circular plate immersed in a Brinkman fluid medium.
The solution proceeds through the formulation of the mixed boundary value problem at hand in terms of dual integral equations that are eventually transformed into Fredholm integral equations of the first kind.
We have shown that the kernel functions can be expressed in terms of infinite integrals over the wavenumber, subsequently recast in the form of fast converging series expansions.
By solving the resulting Fredholm integral equations and performing the relevant integrations numerically, the Brinkman flow fields can be computed in the whole fluid domain.
Unlike computational fluid dynamics models which generally require considerable
time and memory usage, the present approach shows its robustness and potential advantage in solving fluid mechanics problems based on one dimensional integration only.
In the limit of infinite plate radius or zero impermeability, our solution is found to be in full agreement with previous studies, confirming the validity and reliability of our approach.
More importantly, we show that the effect of the plate on the swimming behavior of a self-propelling dipole swimmer is maximum when the radius of the plate is comparable to the distance separating the swimmer from the plate.

While the dual integral equations approach developed in the present work is exclusively valid for an axisymmetric flow, it can be extended to be applicable for an asymmetric flow situation as well.
For an arbitrary position of the point force above the plate, one can follow an alternative route based on the solution strategy described by Miyazaki~\cite{miyazaki84}.
The latter made use of the Green and Neumann functions supplemented by the edge function to remove the singularity at the rim of the disk to obtain closed-form solutions of the hydrodynamics equations for a Stokeslet singularity acting near a no-slip disk in the Stokes limit.
A general solution of the Brinkman problem near a plate is of relevance and is worth investigation following Miyazaki's approach in a future work.

We believe that the solution derived in the present work may find applications in the context of microfluidics, for instance, in the design and control of artificial self-propelling microrobots in a Brinkman fluid medium.
One outstanding example is a low-Reynolds-number swimmer consisting of two coaxially positioned circular disks intercalated by a spherical particle of small size.
These three elements are connected by rod-like constituents of negligible hydrodynamic effects in order to ensure their axial alignment.
Self propulsion is achieved by changing the mutual distance between the elements in a non-reciprocal manner such that the time-reversal symmetry of Stokes flow is broken; see Ref.~\cite{nickandish2021dynamic} where an analogous design has been proposed recently.
Based on the solution obtained here, the overall behavior of the swimmer can be fully analyzed and characterized in terms of the underlying physical and geometrical properties of the system.
While in many practically relevant situations moving to a computational fluid dynamics solver could provide more flexibility, the present semi-analytical approach may prove useful in making approximate estimates of key system properties such as the hydrodynamic reaction tensor without recourse to expensive numerical simulations.

\begin{acknowledgements}
We acknowledge support from the Max Planck Center Twente for Complex Fluid Dynamics, the Max Planck School Matter to Life, and the MaxSynBio Consortium, which are jointly funded by the Federal Ministry of Education and Research (BMBF) of Germany and the Max Planck Society. This work was supported by Slovenian Research Agency (A.V., grant number P1-0099).
\end{acknowledgements}

\appendix

\section{Expressions of the series coefficients}
\label{appendix:coeffs}

In this Appendix, we provide the expressions of the series coefficients defining $\Psi_i$, $i = 1, \dots, 4$, given by Eqs.~\eqref{Psi1_Psi2} and~\eqref{Psi3_Psi4}.
The expressions of the series coefficients $X_m$, $C_m$, $T_m$, and~$U_m$ involve Gauss (or ordinary) hypergeometric function and are given in Tab.~\ref{tab:coeffs1}.
The expression of the series coefficients $S_m$, $V_m$, $Q_m$, and~$W_m$ involve a generalized hypergeometric function and are given in Tab.~\ref{tab:coeffs2}.
The remaining series coefficients are expressed as 
\begin{align}
    Z_m &=  \frac{\pi}{2} \, C_m - \frac{4}{3} \, m\mu S_m
	- \frac{2}{\mu} \frac{Q_m}{2m+1} \, , \label{Z_m} \\
    G_m &=  \pi \, T_m - 2(2m+1)  \mu  V_m  - \frac{1}{\mu} \frac{W_m}{m+1} . \label{G_m}
\end{align}

\bgroup
\def\arraystretch{3} 
    \begin{table}
    \centering
    \begin{tabular}{|c|rrrrrrr|}
    \hline
    ~$X_m$~  & ~$ {}_2F_1$ & $ \bigg($ & $ -m,$ & $\displaystyle -\frac{1}{2}-m ;$ & 2; &  $\mu^2$ &  $\bigg)$~ \\[8pt]
    \hline
    ~$C_m$~ &  ~$ {}_2F_1$ & $ \bigg($ & $ -m,$ & $\displaystyle \frac{1}{2}-m; $ & 2; & $\mu^2$ &  $\bigg)$~ \\[8pt]
    \hline
    ~$T_m$~ & ~$ {}_2F_1$ & $ \bigg($ &  $-m,$ & $\displaystyle -\frac{1}{2}-m; $ & 1; & $\mu^2$ & $\bigg)$~ \\[8pt]
    \hline
    ~$U_m$~ & ~$ {}_2F_1 $ & $ \bigg($ &  $-1-m,$ & $\displaystyle -\frac{1}{2}-m; $ & 1; & $\mu^2$ & $\bigg)$~ \\[8pt]
    \hline
    \end{tabular}
        \caption{Expressions of the series coefficients expressed in terms of Gauss hypergeometric function~${}_2F_1$.}
        \label{tab:coeffs1}
    \end{table}
\egroup

\bgroup
\def\arraystretch{3} 
    \begin{table}
    \centering
    \begin{tabular}{|c|rrrrrrrrr|}
    \hline
    ~$S_m$~ &  ~$ {}_3F_2 $ & $\bigg($ &  1, & $1-m,$ & $\displaystyle \frac{1}{2}-m;$ & $\displaystyle \frac{3}{2},$ &  $\displaystyle \frac{5}{2};$ & $\mu^2 $ & $\bigg)$~ \\[8pt]
    \hline
    ~$V_m$~ &  ~$ {}_3F_2$ & $ \bigg($ &  1,& $-m,$ & $\displaystyle \frac{1}{2}-m;$ & $ \displaystyle\frac{3}{2},$ & $\displaystyle\frac{3}{2};$ & $ \mu^2 $ & $\bigg)$~ \\[8pt]
    \hline
    ~$Q_m$~ &  ~$ {}_3F_2 $ & $\bigg($ & $\displaystyle -\frac{1}{2},$ &  $\displaystyle \frac{1}{2},$ &  1; &  $m+1,$ &  $\displaystyle m+\frac{3}{2};$ & $\displaystyle \frac{1}{\mu^{2}}$ &  $\bigg)$~ \\[8pt]
    \hline
    ~$W_m$~ & ~$ {}_3F_2 $ & $\bigg($ &  $\displaystyle \frac{1}{2},$ &  $\displaystyle \frac{1}{2},$ & 1; &  $m+2,$ & $\displaystyle m + \frac{3}{2} ;$ &  $\displaystyle \frac{1}{\mu^{2}}$ & $\bigg)$~ \\[8pt]
    \hline
\end{tabular}
        \caption{Expressions of the series coefficients expressed in terms of the generalized hypergeometric function~${}_3F_2$.}
        \label{tab:coeffs2}
    \end{table}
\egroup

\section{Truncation of infinite series}
\label{appendix:trunc}

As outlined in the main body of the paper, the kernel functions can be cast in the form given by Eq.~\eqref{eq:kernels_final}.
On the one hand, the kernel $\Gamma_1$ involve the functions $\Psi_1$ and $\Psi_2$ which are expressed in series forms by Eqs.~\eqref{Psi1_Psi2}.
$\Psi_1$ and $\Psi_2$ are given in terms of the coefficients $X_m$, $C_m$, and $Z_m$.
On the other hand, the kernel $\Gamma_2$ involve the functions $\Psi_3$ and $\Psi_4$ which are likewise expressed in series forms by Eqs.~\eqref{Psi3_Psi4}.
$\Psi_3$ and $\Psi_4$ are given in terms of the coefficients $T_m$, $U_m$, and $G_m$.
Meanwhile, $Z_m$ and $G_m$ are expressed as a combination of some of these series coefficients and are given by Eqs.~\eqref{Z_m} and \eqref{G_m} of Appendix~\ref{appendix:coeffs}.

The series representation of the kernel functions offer a great computational advantage over the direct numerical evaluation of the corresponding infinite integrals.
The number of terms required to achieve a given desired precision overall depends on the magnitude of $\sigma = \alpha t/2$ and~$\mu = r/t$.
Generally, only a few terms are needed to achieve good precision. 
However, as these parameters get larger, a large number of terms is typically needed for an accurate computation of the infinite series.

To truncate the infinite series up to a certain number or terms, it is useful to examine the behavior of the general term of the series at infinity.
Since the series coefficients are defined in terms of hypergeometric functions, probing the asymptotic behavior of the general terms of the series for arbitrary values of $\mu$ is delicate and far from being trivial.
We thus restrict the following discussion in the limit $\mu \gg 1$ for which a large number of terms is generally required for an accurate computation of the series.

Defining $P_m = \mu^{2m}/ \left( \pi m \right)^\frac{1}{2}$, 
in the limits when $\mu \to \infty$ and~$m \to \infty$, it can be shown that
$X_m \sim 2 P_m$, 
$C_m \sim P_m / m$, 
$S_m \sim 3\pi P_m / \left( 4m \mu^2\right)$, 
$T_m \sim 2m P_m$,
$U_m \sim \mu^2 P_m$,
and $V_m \sim \pi P_m / \left( 4m \right)$.
In addition, both $Q_m$ and~$W_m$ are found to be of order one to leading order.
We also have $\Gamma(m+a) \sim \left( 2\pi\right)^\frac{1}{2} m^{m+a-\frac{1}{2}} e^{-m}$, $a \in \mathbb{C}$, in the limit $m \to \infty$.
It follows from Eqs.~\eqref{Z_m} and~\eqref{G_m} that $Z_m \sim \pi P_m / \left( 2m \right)$ and $G_m \sim -\pi \mu P_m$.
Accordingly, the general term of the series defining $\Psi_1$ is approximately given by $C_{11} \left( \mu \sigma e/m\right)^{2m} m ^{-\frac{7}{2}}$ for $\sigma \ne 1/2$, and by $C_{12} \left( \mu e/m/2\right)^{2m} m ^{-\frac{9}{2}}$ for $\sigma = 1/2$, where $C_{11} = \left( 2\sigma-1\right) / 8 / \pi^{\frac{1}{2}}$ and $C_{12} = - 5 /32 /\pi^\frac{1}{2}$.
In addition,  the general term of the series defining $\Psi_2$ is given by $C_2 \left( \mu \sigma e/m\right)^{2m}m ^{-\frac{7}{2}}$, where $C_2 = 1/16/\pi^\frac{1}{2}$.
Finally, the general terms defining  $\Psi_3$ and~$\Psi_4$ are approximately given by $ C_3 \left( \mu \sigma e/m\right)^{2m}m ^{-\frac{3}{2}}$ and $C_4\left( \mu \sigma e/m\right)^{2m}m ^{-\frac{5}{2}}$, respectively, where $C_3 = 1/\pi^\frac{1}{2}$ and $C_4 = 1/4/\pi^\frac{1}{2}$.

Defining $\upsilon = \mu \sigma e/M$, where $M$ is integer at which the series is truncated, it follows that, for $0 < \upsilon < 1$, the truncation errors can be cast in the form
\begin{equation}
    \mathcal{E}_M \simeq \left| \sum_{m=M}^\infty C \upsilon^{2m} m^{-\zeta} \right| \, , 
\end{equation}
where $\zeta \in \{3/2,5/2,7/2,9/2 \}$.
Then, 
\begin{equation}
    \mathcal{E}_M < \sum_{m=M}^\infty |C| \upsilon^{2m} = |C| \frac{\upsilon^{2M}}{1-\upsilon} < \epsilon \, , \label{eq:error}
\end{equation}
where $\epsilon$ is the desired truncation error.
Solving Eq.~\eqref{eq:error} for~$M$ provide an estimate of the number of terms required to achieve a certain precision when $\mu \gg 1$.
For $\mu$ and~$\sigma$ less than unity, only a very few number of terms is generally required.

During the numerical computation of the series, we check after incrementing the summation index that the \textit{absolute approximate error} is less than desired truncation error by ensuring that the absolute value of the ratio between the $M$th summation term and the approximate sum is less than~$\epsilon$.
Specifically, denoting by $f_n$ the general term of a given series, then the series is truncated after the condition
\begin{equation}
    \left| \frac{f_M}{\sum_{m=0}^M f_m } \right| <  \epsilon \label{eq:error_num}
\end{equation}
is satisfied.

\bgroup
\def\arraystretch{1.5} 
    \begin{table}
    \centering
\begin{tabular}{c|cccc|cccc|cccc|cccc}
     \multirow{2}{*}{$\epsilon$} & \multicolumn{4}{c|}{$\Psi_1$} & \multicolumn{4}{c|}{$\Psi_2$} & \multicolumn{4}{c|}{$\Psi_3$} & \multicolumn{4}{c}{$\Psi_4$} \\
    \cline{2-17}
     & A&B&C&D & A&B&C&D & A&B&C&D & A&B&C&D \\
    \hline
    \hline
    $10^{-3}$ & 7&5&33&41 & 4&1&21&27 & 9&6&34&42 & 4&1&23&29 \\
    \hline
    $10^{-6}$ & 10&8&37&45 & 10&3&29&34 & 12&9&38&46 & 9&3&29&34 \\
    \hline
    $10^{-9}$ & 13&10&41&48 & 12&6&34&39 & 14&11&42&50 & 12&7&34&39 \\
    \hline
\end{tabular}
\caption{Number of terms required for the computation of the infinite series defining $\Psi_i$, $i=1, \dots, 4$, for three truncation errors and four sets of parameters $(\sigma, \mu)$; see main text for the corresponding values.}
        \label{tab:error}
    \end{table}
\egroup

Table~\ref{tab:error} provides the number of terms required for the computation of the series functions defining $\Psi_i$, $i=1, \dots, 4$, based on the stopping criterion given by Eq.~\eqref{eq:error_num}, for three values of the truncation error.
Four exemplary situations are shown for the set of parameters $(\sigma, \mu)$ given by $(2,1)$, $(1,2)$, $(3,5)$, and~$(5,3)$ corresponding to the cases A, B, C, and~D, respectively.
The series exhibit fast convergence behavior where only a few additional terms are typically required to improve the accuracy by orders of magnitude.
We further notice that a smaller number of terms is typically required for the computation of~$\Psi_2$ and~$\Psi_4$. 
Throughout this work, we have consistently evaluated all the series based on a truncation error of $10^{-10}$.

\section{Flow-field related expressions in the Stokes limit}
\label{appendix:exp}

The expressions of $\mathcal{K}_i$, $i=1, \dots, 4$, stated by Eqs.~\eqref{eq:Ks} take a particularly simpler form in the limit $\lambda \to 0$,
\begin{subequations}
    \begin{align}
    \mathcal{K}_1 &= - \int_0^\infty \left( q|z|-1 \right) e^{-q|z|} \sin(qt) J_1(qr) \, \mathrm{d}q \, , \\
    \mathcal{K}_2 &= -|z| \int_0^\infty q e^{-q|z|} \cos(qt) J_1(qr) \, \mathrm{d}q \, , \\
    \mathcal{K}_3 &= -|z| \int_0^\infty q e^{-q|z|} \sin(qt) J_0(qr) \, \mathrm{d}q \, , \\
    \mathcal{K}_4 &= -\int_0^\infty \left( q|z|+1 \right) e^{-q|z|} \cos(qt) J_0(qr) \, \mathrm{d}q \, .
\end{align}
\end{subequations}
We note that in this limit, we have $\mathcal{Q}_1/\mathcal{K}_3 = 2\eta/|z|$.

%

\end{document}